\documentclass[11pt,a4paper]{article} 
\makeatletter
\let\@fnsymbol\@arabic
\makeatother
\usepackage[T1]{fontenc}
\usepackage[utf8]{inputenc}
\usepackage{authblk}
\usepackage{longtable}
\usepackage[hyphens]{url}
\usepackage{graphicx}
\usepackage{pdflscape}
\usepackage{float}
\usepackage{color}
\usepackage[stable]{footmisc}
\usepackage[english]{babel} 
\usepackage{ifthen}
\usepackage{fancyhdr}
\usepackage{natbib}
\usepackage{url}
\usepackage{subfigure}

\makeatletter
\let\@fnsymbol\@arabic
\makeatother
\title{Matching Entities Across Online Social Networks}
\author{Olga Peled\thanks{olgit23@gmail.com}}
\author{Michael Fire\thanks{mickyfi@bgu.ac.il}}
\author{Lior Rokach\thanks{liorrk@bgu.ac.il}}
\author{Yuval Elovici\thanks{elovici@bgu.ac.il}}
\affil{Telekom Innovation Laboratories  at Ben-Gurion University of the Negev 

}

\date{}

\begin{document}
\maketitle

\begin{abstract}
Online Social Networks (OSNs), such as Facebook and Twitter, have  become an integral part of our daily lives. There are hundreds of OSNs, each with its own focus in that each offers particular services and functionalities. Recent studies show that many OSN users create several accounts on multiple OSNs using the same or different personal information. Collecting all the available data of an individual from several OSNs and fusing it into a single profile can be useful for many purposes. In this paper, we introduce novel machine learning based methods for solving Entity Resolution (ER), a problem for matching user profiles across multiple OSNs. The presented methods are able to match between two user profiles from two different OSNs based on supervised learning techniques, which use features extracted from each one of the user profiles. By using the extracted features and supervised learning techniques, we developed classifiers which can  perform entity matching between two profiles for the following scenarios: (a) matching entities across two OSNs; (b) searching for a user by similar name; and (c) de-anonymizing a user's identity.

The constructed classifiers were tested by using data collected from two popular OSNs, Facebook and Xing. We then evaluated the classifiers' performances using various evaluation measures, such as true and false positive rates, accuracy, and the Area Under the receiver operator Curve (AUC). The constructed classifiers were evaluated and their 
classification performance measured by AUC was quite remarkable, with an AUC of up to 0.982 and an accuracy of up to 95.9\% in 
identifying user profiles across two OSNs.
\\\\
\noindent \textbf{Keywords.} Online Social Networks; Entity Matching; Online Social Network Privacy; Machine Learning.

\end{abstract}

\section{Introduction}
\label{intro}
In the last decade, the Internet has greatly expanded and evolved to become a central part of our daily lives. Activities that used to take place manually, like shopping, playing games, and paying bills are now being implemented online. Today, even a major part of our social lives takes place on the Internet (\cite{veldman2009matching}). In particular, social networking has now become one of the most popular and utilized activities on the web. Social network users have access to a ``second life'' on OSNs, such as Facebook,\footnote{\url{http://www.facebook.com/}} Twitter,\footnote{\url{http://www.twitter.com/}} and LinkedIn.\footnote{\url{http://www.linkedin.com/}} People use OSNs for various purposes, like 
keeping in touch with friends, joining groups, playing online games, and sharing information (\cite{veldman2009matching}). There are hundreds of OSN sites, each network with its own focus and its own particular services and functionalities offered to its users (\cite{veldman2009matching}). 
Facebook, for example, is one of the most popular social networking websites on the Internet with more than 1.23 billion monthly active users as of December 2013 (\cite{fbstats2014}). Facebook is intended to connect friends, family, and business associates. Users create profile pages which present information about themselves as well as their friends list. Moreover, Facebook also allows its members to communicate, connect, and share information with each other by chatting, posting messages and comments on each other's walls, and even liking each other's posts (\cite{wilson2012review}). To make use of the provided services and functionalities, and to keep updated with other members, users tend to create several accounts on various sites where their friends may also be registered to multiple networks (\cite{veldman2009matching}). This phenomenon has been studied in the last several years.
\cite{patriquin2007} reported on the member overlap between various OSNs services. He showed that 64\% of Facebook users also have MySpace\footnote{\url{http://www.myspace.com/}} accounts and 42\% of LinkedIn users also have Facebook accounts. Moreover, according to the recent Pew Research Center's Internet Project survey results, 42\% of the adults online  use two or more social networking platforms,  while 36\% use only one social network platform (\cite{pewinternet2013}).

In most cases, to join different OSNs users must register separately to each one by filling out personal information forms. In other cases, people can register to a new OSN using their account credentials from different networks. For instance, Facebook users can register to Academia.edu\footnote{\url{http://academia.edu/}} through their Facebook account. Users who create several accounts on various sites (e.g., for personal use, for work, etc.), can use the same or different personal information. For example, the personal information on Miley Cyrus's Google+ profile\footnote{\url{https://plus.google.com/+MileyCyrus/about}} says that she lives in
Los Angeles, CA, while her Facebook profile\footnote{\url{https://www.facebook.com/MileyCyrus/info}} shows that her hometown is Franklin, TN. In many cases, although people provide their personal information to multiple OSNs, they do not intend to provide data for integration with other information sources.
Therefore, the integration of personal information can be quite a sensitive issue for privacy reasons. 
Inter-social network operations and functionalities are required for several operations, such as data integration, data enrichment, information retrieval, etc. (\cite{raad2010user}). 

Retrieving all available data from one person and merging it into a single profile can be a useful tool for many purposes, like building a full profile of a real individual, aggregating online friends into a single integrated list, searching for people across different social networks, and even for employers to track their employees' actions on OSN sites (\cite{vosecky2009user}).  Moreover, data integration can be useful for a company who wishes to enrich current customer data to improve marketing (\cite{veldman2009matching}). 
The process of identifying different profiles, which belong to the same real individual, is known as Entity Resolution (ER) (\cite{veldman2009matching}). Today there are various sites, such as Snitch.Name\footnote{\url{http://snitch.name/ }}  and Pipl\footnote{\url{http://pipl/}} that allow for the searching of users across multiple OSN sites by name. However, cross-referencing users between different sites is not an easy task since in many cases, users of different OSNs fill in different details, use different user names, and have different lists of friends. To complicate the problem even further, many different users have similar names and personal details. For example, according to a frequency table of names on Facebook,\footnote{\url{http://www.skullsecurity.org/blog/2010/return-of-the-facebook-snatchers}} 17,204 out of 100 million people are named John Smith. As a result, even if two user profiles have the same first and last name, it is not enough to confirm that these two user profiles belong to the same real person. 
In this study, we present an algorithm for solving the ER problem for profiles from different OSNs. We attempt to find out which user profiles of the first social network correspond to user profiles of the second social network, where we assume a person to have at most one profile per OSN site. Our algorithm performs entity matching between two user profiles from two different OSNs for three main scenarios: (a) \textit{matching users across two social networks} - given two user profiles from different social networks, we calculate the probability that these two user profiles belong to the same real individual; (b) \textit{searching for a user} - in this scenario we simulate a case in which we have a person's name, and we would like to find his or her accounts, which contain a similar name, on different social networks; and (c) \textit{de-anonymizing user's identity} - in this scenario we simulate a case in which we have a person's name and profile details in one social network, and we would like to find his or her hidden accounts, which contain a pseudonym, on different social networks. Our solution is based on machine learning techniques that use feature extraction on a user's information and on its friends list. We only use public user profiles, therefore we are not violating users' privacy. Different classifiers are trained and used for each scenario. These classifiers utilize multiple features of mainly three types: (a) name based features, such as the Soundex (\cite{holmes2002improving,lait1996assessment}) value of two names and Edit Distance (\cite{navarro2001guided}) of two names; (b) user information-based features, such as the cosine (\cite{cohen1998integration}) and Jaccard similarity (\cite{jaccard1901etude}) between two full user profiles; and (c) social network topological based features, such as the number of mutual friends (\cite{liben2007link}) between the two users' friends list. In contrast to previous studies, our algorithm uses machine learning techniques with feature extractions of many users' data features, like name, birthday, location, professional experience, education, etc. Additionally, previous studies evaluated their methods on small datasets containing about 3,000 users (\cite{vosecky2009user}), while in this study we evaluated our ER algorithm on large datasets, which included more than 30,000 users obtained from two different popular OSNs, Facebook and Xing.\footnote{\url{http://www.xing.com/}} 

This experimental study uses real-life data collected from two popular OSNs, Facebook and Xing. The proposed algorithm was evaluated, and its classification performance measured by AUC was 0.982 in identifying user profiles across two OSNs. In addition, the classification performance measured by AUC was 0.88 for the second scenario in which we utilize a user's details in one social network to identify his or her profile, with a similar name, in another social network. Moreover, the classification performance measured by AUC was 0.86 for the third scenario, in which we simulated a case in which a person's name and profile details in one social network are given and we would like to identify his or her hidden accounts, which contain a pseudonym on different social networks. The main contribution of our matching algorithm is that we use machine learning techniques and suggest many new features. Moreover, we demonstrate that our presented algorithms can also match users across two social networks without utilizing the users' name features.

The remainder of the article is organized as follows. In Section~\ref{sec:related}, we provide a brief overview of previous studies on entity resolution, particularly in social networks. We also describe several string matching methods and techniques. In Section~\ref{sec:problem}, we describe and define the problem formulation. In Section~\ref{sec:method}, we describe the methods used for developing and evaluating entity resolution on OSNs.  The features used in this study are formally defined in Section~\ref{sec:features}. Section~\ref{sec:exper} describes our experimental study and provides the results obtained by running the algorithm on real world OSNs. In this section, we also present our obtained results, including the AUC, classification accuracy, and the contribution of each feature set. In Section~\ref{sec:disc}, we discuss the obtained results. Lastly, in Section~\ref{sec:conclusion}, we present our conclusions and point out future research directions.

\section{Related Work}
\label{sec:related}
Entity Resolution (ER) is an important issue, which has been addressed by many researchers in the past (\cite{getoor2005link,brizan2006survey,elmagarmid2007duplicate,benjelloun2009swoosh}). In our study we address a very specific case of ER, namely, the matching of user profiles across multiple OSNs. In this section we describe various related research results and discuss their connection to our problem. We begin with the definition of social networks (see Section~\ref{sec:related_sn}) and continue with the current approaches concerning ER, particularly in social networks (see Section~\ref{sec:related_er}). We conclude this section with the description of several string matching methods and techniques (see Section~\ref{sec:related_sm}).

\subsection{Social Networks}
\label{sec:related_sn}
\cite{ellison2007} define social network sites as ``web-based services that allow individuals to: 
\begin{itemize}
\item construct a public or semi-public profile within a bounded system. 
\item articulate a list of other users with whom they share a connection. 
\item view and traverse their list of connections and those made by others within the system.'' 
\end{itemize}

Social networking is an online service and platform that focuses on facilitating the building of OSNs among people who, for example, share interests, activities, backgrounds or real-life connections. OSNs allow users to share ideas, activities, events, and interests within their individual networks. People use OSN sites to meet new people and to communicate with people who are already a part of their life (\cite{ellison2007}). Social networks have a representation of each user profile that describes the user's information, such as  age, location, interests, pictures, work and education, social links (friends), as well as a variety of additional services, such as e-mail, instant messaging, and photo-sharing. The visibility of a profile varies by site and according to the individual user's discretion. Furthermore, structural variations around visibility and access are one of the primary ways that social networks differentiate themselves from one another (\cite{ellison2007}). Some sites are designed with specific ethnic, religious, sexual orientation, political or other identity-driven categories in mind. There are even social networks for pets, such as Dogster\footnote{\url{http://www.dogster.com/}} and Catster,\footnote{\url{http://www.catster.com/}} where the pet owners manage their pets' profiles. 

\subsection{Entity Resolution}
\label{sec:related_er}
ER addresses the issue of duplication and uses algorithms to both detect duplicates in data and to resolve them.  According to \cite{veldman2009matching}, the ER solution can be divided into three different groups: supervised, unsupervised, and semi-supervised solutions. Supervised approaches need labeled training datasets or predefined thresholds to base their decisions on. Unsupervised approaches avoid human intervention by using clustering algorithms that group together items which present a high similarity. The semi-supervised approach uses a small set of labeled data and a set of unlabeled data. The training phase is then skipped, and together with the labeled data, the unlabeled data is resolved.
In this study we present a new supervised approach in order to resolve the ER problem. 
In the following subsections we broadly describe the supervised and the unsupervised approaches. 

\subsubsection{Supervised Approaches}
The supervised approaches rely on the existence of training data in the form of record pairs, pre-labeled as match or not match. \cite{fellegi1969theory} described the ER problem as a simple problem in which there are two classes: a class M, which represents matches, and a class U, which represents non-matches. Each pair needs to be assigned to one of these classes. There are three main approaches (\cite{elmagarmid2007duplicate,veldman2009matching}):

\paragraph{Learning-based.}  Learning-based solutions use training datasets which consist of pairs labeled as match or non-match. Each pair has a comparison vector that represents the comparable attributes of the two items in the pair. For the classes M and U, the assumption is that the density functions are different. Now, the problem of ER can be treated as a Bayesian inference problem.

\paragraph{Distance-based.} Distance-based solutions do not need training data. In these approaches a distance metric is defined between data items. Based on the distance between two items, a decision is made on whether or not this pair is a match using a threshold. This threshold can be set by making a reliable estimation.  A good threshold can improve the results. Therefore, using some kind of training set is recommended.

\paragraph{Rule-based.} Rule-based solutions base decisions on rules from domain knowledge. For example, if two items contain the same email, you can conclude that the items refer to the same person.

In this study, our approach is mainly learning-based.  For instance, we use training datasets that contain pairs of user profiles, each from a different network and for each pair there is a set of features and a label (match or not match). 

In all the approaches mentioned above, decisions are made for each pair in isolation, which is a more classical approach. The recent research trend is to eliminate the naive assumption that these decisions are independent.

\subsubsection{Unsupervised Approaches}
The idea behind unsupervised approaches is that similar comparison vectors correspond to the same class (\cite{elmagarmid2007duplicate}). The idea of unsupervised learning for duplicate detection has its roots in the probabilistic model proposed by \cite{fellegi1969theory}. When there is no training data to compute the probability estimates, it is possible to use variations of the Expectation Maximization (\cite{dempster1977maximum}) algorithm to identify appropriate clusters in the data.

\subsubsection{Entity Resolution Across Different Social Networks}
Our study deals with entity matching in OSNs, thus, in this section we will describe the main approaches to match user profiles across multiple OSNs that belong to the same real individual. 
The tool D-Dupe, presented by~\cite{bilgic2006d}, uses networks to perform ER. In the domain of scientific publications Bilgic et al. used the co-authorship relation to detect duplicates. For each author in their dataset, they determined the network of co-authors with whom this author has written an article. In this network of co-authors, if two authors in the network have very similar names and the titles of the articles they wrote are also very similar, then these authors are probably the same person. This approach differs from ours in that they use one source for which duplicates are present. Since the duplicates appear within one network, similarities between co-authors or their associated information, such as article, title or journal, provide a very probable hint that these co-authors are in fact the same person. Moreover, this tool leaves the user to decide whether or not two authors match, as it can only provide hints. For small sets, this is a feasible and efficient resource. 

As many users have profiles on multiple OSNs, there is a need for linking techniques. From the user's perspective, there is a need for so-called ``social network aggregation,'' which provides a mechanism for maintaining multiple profiles from one interface. These aggregators are not concerned with the ER problem themselves, since the requests for integration of the profiles comes from the user. Examples of some aggregators are AOL Lifestream\footnote{\url{http://lifestream.aol.com/}} and FriendFeed.\footnote{\url{http://friendfeed.com/}} Although they appear to solve the ER problem of multiple user profile pages, these links only work if a user is willing to cooperate. In addition to these initiatives, there are search engines like Pipl that try to collect as much personal information as possible from social networks about a particular person. However, these sites can only match names, which is too restrictive. 

\cite{vosecky2009user} presented a matching technique in which each user profile is represented as a vector consisting of the values of individual profile fields (e.g., name, date of birth, etc.). The comparison between any two vectors consists of two phases. In the first phase, the algorithm calculates a similarity score between corresponding vector fields using an appropriate string matching function for each field, resulting in a similarity vector. In the second phase, a weighting vector is applied to the similarity vector to calculate the overall similarity. This method was trained to determine an optimal set of profile fields and tested on less than 3,000 users from Facebook and StudiVZ\footnote{\url{http://www.studivz.net}} (a German language social network), with fifty mutual users (the database was manually checked). The result had about an 83\% accuracy rate in identifying duplicated users across two social networks.

\cite{raad2010user} represented each user profile as a FOAF file.  FOAF is a machine-readable ontology describing people, their activities, and their relations to other people. The comparison between any two FOAF files consists of three phases. In the first phase, a weight is assigned, manually or automatically to each attribute in the FOAF file. Then, string and semantic similarity metrics are used to compare attribute values. Lastly, aggregation functions are used to compare attribute values.  In the last phase, aggregation functions are used for data fusion and for decision making. This technique was tested on an artificial dataset only. 

\cite{veldman2009matching} suggested two models to solve the ER problem across two social networks. In the simple model, Veldman compared all the profiles of the first network against all the profiles of the second.  Each compared pair received a pairwise similarity score. The higher this score was, the higher the probability was that these profiles belonged to the same real person. The pairs that satisfied the so-called pairwise threshold were the candidate matches, and from these candidate matches the final matches were chosen. In addition to the simple model, Veldman also used the network model in that for each candidate match, the network similarity score is calculated (i.e., friends). This is done by determining the overlap in the networks of both profiles of the candidate match. The more the networks overlap, the higher the network similarity score, and the higher the probability that the two user profiles belong to the same real person. Again, the candidate matches should satisfy a network threshold to remain as such. Then, from the remaining candidate matches, the final matches can be chosen. 

\cite{carmagnola2010user} presented the preliminary results of their research work aimed at uniquely identifying users on different social systems and retrieving their data distributed over the profiles stored in such systems. Their proposed algorithm requires some initial attributes about the user to set up the search. Given a set of input attributes about said user, crawlers search for user profiles that match the input request. Once the user profile has been retrieved, with a matching nickname or full name, the other initial attributes are used to compute a score for this match. Each attribute has its own weight based on how strong a positive match indicator it is. Lastly, a probability is calculated which represents the probability that the newly discovered user attributes actually belong to the searched user.  Even though it offers promising results, this approach needs further advancement before it will be fully functional and productive.

\cite{iofciu2011identifying} went beyond user matching based on profile information and included content-based information (i.e., tags that users assigned to images and bookmarks), when matching Flickr,\footnote{\url{http://www.flickr.com/}} Delicious,\footnote{\url{ http://delicious.com/}} and StumbleUpon\footnote{\url{http://www.stumbleupon.com/}} user accounts. They suggested combining profile attributes (e.g., usernames), with an analysis of the tags contributed by them to identify users. They also suggested various strategies to compare the tag-based profiles of two users. 
Their experiments achieved an accuracy of almost 80\% in user identification.

\cite{carmagnola2009user} developed a framework that provides a common base for user identification for cross-system personalization among web based user adaptive systems. Their approach bases heuristics on profile attributes, such as username, name, location or the email address of the user. 
Narayanan and Shmatikov illustrated in two recent papers that information from different data sources can be combined in order to identify a user (\cite{narayanan2008robust,narayanan2009anonymizing}). They further showed that statistical methods can be applied to identify micro-data by cross-correlating multiple datasets (\cite{narayanan2008robust}). They extended their approach to social networks and proved that it is possible to identify members by mapping known, auxiliary information on the social network topology (\cite{narayanan2009anonymizing}). Narayanan and Shmatikov also demonstrated that a third of users with accounts on both Twitter and Flicker can be re-identified in an anonymous Twitter graph with only a 12\% error rate.
In our study, just as in previous work, we considered a range of profile fields and investigated their importance for the matching process. However, unlike previous studies, we used supervised machine learning techniques and constructed classifiers that utilized a vast variety of features.  Moreover, we tested our methods on a real and large dataset that contained over 30,000 entities.

\subsection{Distance Metrics}
\label{sec:related_sm}
One of the most common sources of mismatches in database entries is the typographical variations of string data. To deal with typographical variations, duplicate detection typically relies on string comparison techniques.
In this section, we describe string matching techniques that have been applied in duplicate record detection context.

\subsubsection{Character-Based Similarity Metrics}
The character-based similarity metrics are designed to efficiently handle typographical errors. One particular character based method is the Levenshtein distance (also referred to as the edit-distance) (\cite{navarro2001guided}).  This method measures the minimal number of insertions, deletions or substitutions that are needed to transform one string into another string.
The Damerau-Levenshtein distance (\cite{damerau1964technique,navarro2001guided}) is a variation of the edit distance where a transposition of two characters is also considered to be an elementary edit operation.
The Jaro distance (\cite{yancey2005evaluating}) metric is calculated using the number of common characters (i.e., the same characters that are within half the length of the longer string) and the number of transpositions (i.e., swapping of two letters).
Winkler (\cite{yancey2005evaluating}) (or Jaro-Winkler) is a variant of the Jaro distance and has a higher weighing factor for prefixes. Thus, giving higher scores to prefixes could help match abbreviations.

\subsubsection{Token-Based Similarity Metrics}
Character-based similarity metrics fail to capture the similarity in cases where typographical conventions lead to a rearrangement of words (e.g., ``John Smith'' versus ``Smith, John''). Token-based metrics try to compensate for this problem. Token-based methods measure the number of matching tokens between two sets of tokens.

The Jaccard measure (\cite{jaccard1901etude}) is the simplest approach. It measures the ratio of equal tokens in the union of tokens of both strings. The disadvantage of this method, however, is that every word has equal weight.
The TFIDF (Term Frequency/Inverted Document Frequency)  is a method that comes from the field of information retrieval (\cite{cohen1998integration}). This method measures the frequency of a term but also corrects this with the importance of the token. This means that common tokens like ``a,'' ``the,'' and ``but'' receive a lower score because they are not discriminative enough.
The cosine similarity (\cite{cohen1998integration}) expresses strings as term vectors, with each word being a dimension in the vector which counts the frequency of this word. The cosine similarity then measures the angle between the vectors, which is a measure for the similarity between the strings.
Unfortunately, these methods do not take misspellings into account, which means that misspellings can incorrectly decrease the similarity score. Hybrid methods can benefit from both the token-based methods as well as the character-based methods.
The token-based method will calculate a score based on the number of similar tokens, and the character-based method determines whether two tokens are similar, or at least similar enough.

\subsubsection{Phonetic Similarity Metrics}
Character-level and token-based similarity metrics focus on the string-based representation of database records. However, strings may be phonetically similar even if they are not similar at a character or token level. For example, the word ``Kageonne'' is phonetically similar to ``Cajun'' despite the fact that the string representations are very different. The phonetic similarity metrics try to address such issues and match such strings.
Soundex (\cite{holmes2002improving,lait1996assessment}), for example, is the best known phonetic encoding algorithm. It keeps the first letter and converts the rest of the string into numbers according to an encoding table.
Additional string comparison methods can be found in a thorough survey written by \cite{christen2006comparison}.

\section{Problem Formulations}
\label{sec:problem}

Let $S_1$ and $S_2$ represent two OSNs, where each OSN consists of many user profile pages which may be interconnected. 
This study's goal is to predict the probability that two given public profiles belong to the same real life entity. 
Formally, given a list of public users profiles 
$ <p_1,p_2, \cdots, p_n> $ in $S_1$, and a list of public users profiles $<\hat{p}_1,\hat{p}_2, \cdots, \hat{p}_m>$ in $S_2$, we calculate the probability that $p_i$ and $\hat{p}_j$ belong to the same real life entity, where our assumption is that a real person has only one user profile page per OSN site.

\section{Proposed Method}
\label{sec:method}
\begin{figure}
\begin{center}
\includegraphics[width=\textwidth]{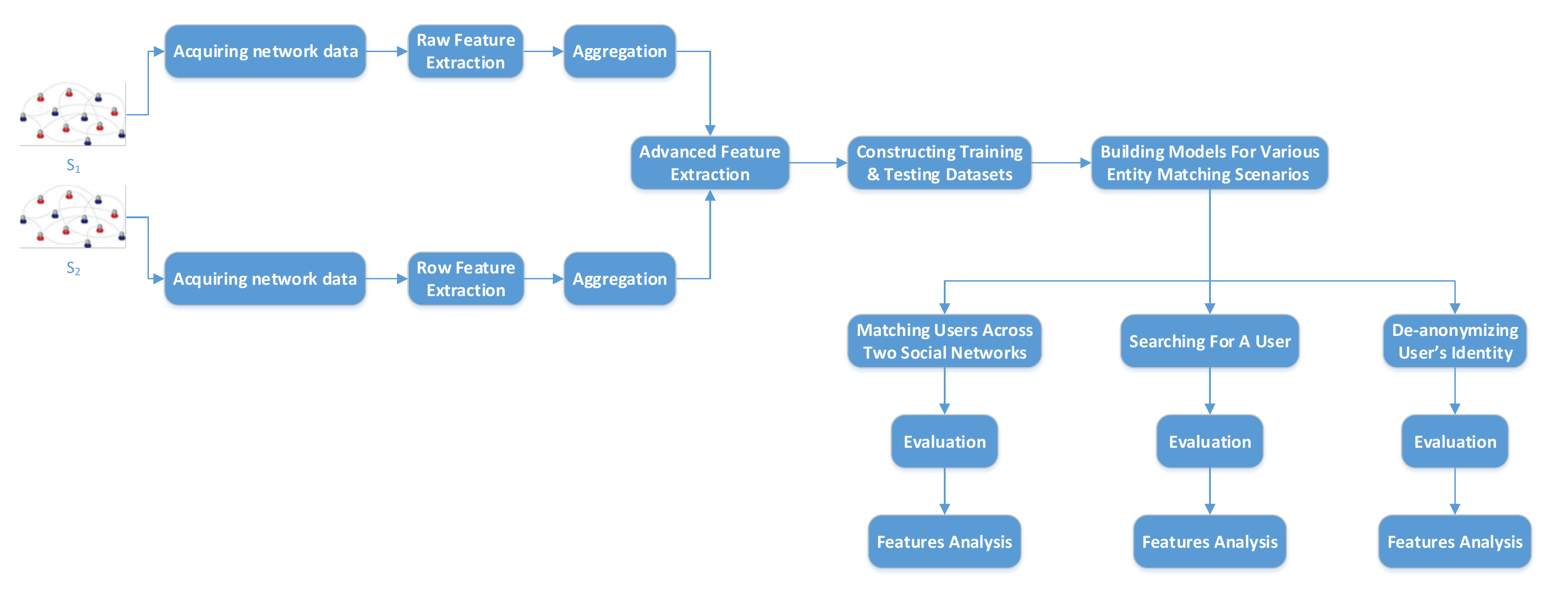}
\end{center}
\caption{Entity resolution process overview}
\label{fig:overview2}       
\end{figure}

We propose a method composed of nine main components, as illustrated in Fig.~\ref{fig:overview2}. Each component is detailed in the following subsections. 

\subsection{Acquiring the Data}
To match user profiles from different OSN sites, a large and suitable dataset from social networks is required. However, it is very difficult to generate such a dataset and the only way to get a realistic dataset is to use real data. Suppose $S_1$ and $S_2$ are OSN sites where people have profile pages on which they share personal information and lists of friends. The data on these profile pages is retrieved using a web crawler. In order to gain access to the public profiles on each of the social networks, we have to use a real account for each network.

\subsection{Raw Features Extraction}
The data on social networking sites can be very diverse, unstructured, and even unsuitable, thus it will need preprocessing. For example, based on the knowledge of the structure of the user profile page and the user's friends' page in a particular network, content can be extracted. Then, irrelevant data can be filtered out. Relevant and non-relevant data depends on the purpose.  After preprocessing, the datasets are ready for the matching process. 

\subsection{Advanced Features Extraction}
\label{sec:features}
After preprocessing, the datasets are ready for the feature extraction process. We extracted 27 features which we compared to two profiles from different OSN sites. We calculated three main types of features: (a) \textit{name based - } features such as the Soundex value (\cite{holmes2002improving,lait1996assessment}), and the Edit distance of two names (\cite{navarro2001guided}); (b) \textit{user information-based - } features such as the cosine similarity between two user profiles (\cite{cohen1998integration}); and (c) \textit{social network topological based - } features such as the number of mutual friends between the two users' friends lists (\cite{liben2007link}). 
Namely, given two user profiles, we calculated the similarity between the different data fields of each user by calculating the following 27  features.


\subsubsection{Name Based Features}
\label{sec:named_based}
Name based features are features that represent the similarity between two names. In this study, we extracted the following 10 name based features for any two given profiles: 
\begin{itemize}
\item \textbf{Soundex Name Similarity} - Soundex (\cite{holmes2002improving,lait1996assessment}) is based on English pronunciation and is the best and oldest known phonetic encoding algorithm. It keeps the first letter in a string and converts the rest into numbers according to an encoding table. The final code is the original first letter and three numbers (longer codes are cut-off and shorter codes are extended with zeros). Soundex is a well-known method in ER (\cite{elmagarmid2007duplicate}). For example, $Soundex(\mbox{``Smith''}) = S530$, 
$Soundex(\mbox{``Smythe''}) = S530$. A major drawback to Soundex is that it keeps the first letter, thus any error or variation at the beginning of a name will result in a different Soundex code. Soundex codes of the two users' full names were compared to see how similar the full names sound when they are spoken. If the two Soundex codes are equal, then the similarity score is 1 and 0 otherwise. For example, 
$Soundex(\mbox{``Smith''}) = Soundex (\mbox{``Smythe''}) = S530$, therefore the similarity name score is 1.  

\item \textbf{Difference Name Similarity} - The Difference function performs a Soundex on two strings and returns an integer that represents how similar the Soundex codes are for those strings. The integer returned is the number of characters in the Soundex values that are the same as one another. The return value ranges from 0 to 4, with 0 indicating weak or no similarity and 4 indicating strong similarity or the same values. For example, $Soundex(\mbox{``olga''}) = O420$ and $Soundex(\mbox{``olgit''}) = O423$; notice that there are 3 of the same characters in the SOUNDEX values, therefore the $Difference(\mbox{``olga'', ``olgit''}) = 3$. To normalize the returned integer value to be in the range of 0 to 1, we divided the returned integer by 4. For example, $Difference(\mbox{``olga'', ``olgit''}) = 3$, therefore
the similarity name is $\frac{3}{4}$.

\item \textbf{LCS Name Similarity} - The Longest Common Sub-string (LCS), (\cite{friedman1992tolerating}) repeatedly finds and removes the longest common sub-string in the two compared strings up to a minimum length. A similarity measure can be calculated by dividing the total length of the common sub-strings by the minimum, maximum or average lengths of the two original strings. This algorithm is more suitable for compound names that have words switched (e.g., given and surname). For example, the total length of the common sub-strings of the two name strings ``gail west'' and ``vest abigail'' is 7. We used the LCS algorithm to calculate the similarity between the two users' full names. After comparing all three ways of calculating the final similarity, we decided to divide the total length of the common sub-string by the average lengths of the two original strings. For example, the $LCS(\mbox{``gail west'',``vest abigail''}) = \frac{7}{10.5}=0.666.	$

\item \textbf{Compression Name Similarity} - The basic idea of Compression Based Similarity (\cite{cilibrasi2005clustering}) is to use the length of the compressed strings (using a standard compressor software like Zip\footnote{\url{http://www.winzip.com}}), to calculate a similarity measure. We used this technique to calculate the similarity score between the two users' full names.

\item \textbf{Damerau Levenshtein Name Similarity} -
Levenshtein, or Edit distance (\cite{navarro2001guided}), is defined to be the smallest number of edit operations, inserts, deletes, and substitutions required to change one string into another. It is calculated using a dynamic programming algorithm. For example, the Edit distance of ``kitten'' and ``sitting'' is 3 (kitten $\rightarrow$ sitten, sitten $\rightarrow$ sittin, sittin $\rightarrow$ sitting).  The distance can be converted into a similarity measure (between 0.0 and 1.0), by dividing it by the maximum length of the two strings. The Damerau-Levenshtein distance (\cite{navarro2001guided,damerau1964technique})  is a variation of edit distance where a transposition of two characters is also considered as an elementary edit operation (in the Levenshtein distance, a transposition corresponds to two edits: one insert and one delete or two substitutions). These two methods are well-known in ER (\cite{elmagarmid2007duplicate}).We used this metric to calculate the similarity between the two users' full names. 

\item \textbf{Jaro-Winkler Name Similarity} - Jaro (\cite{yancey2005evaluating}) is an algorithm commonly used for name matching in data linkage systems (\cite{winkler2006overview}). A similarity measure is calculated using the number of common characters (i.e., the same characters that are within half the length of the longer string), and the number of transpositions. Winkler (or Jaro-Winkler), improves upon the Jaro algorithm by applying ideas based on empirical studies, which found that fewer errors typically occur at the beginning of names {(\cite{yancey2005evaluating})}. The algorithm increases the Jaro similarity measure for up to four agreeing initial characters. This method is well-known in ER (\cite{elmagarmid2007duplicate}). We used this metric to calculate the similarity between the two users' full names. 

\item \textbf{N-Gram Name Similarity} - N-grams are sub-strings of a length n. An n-gram similarity  between two strings is calculated by counting the number of n-grams in common (i.e., n-grams contained in both strings), and dividing the number of n-grams in common by either the number of n-grams in the shorter string (called the Overlap coefficient\footnote{\url{http://simmetrics.sourceforge.net/}}), the number of n-grams in the longer string (called the Jaccard similarity (\cite{jaccard1901etude})), or the average number of n-grams in both strings (\cite{kukich1992techniques}). This method is well-known in ER (\cite{elmagarmid2007duplicate}). We used \textbf{2-grams} and \textbf{3-grams} to calculate the similarity between the two users' full names.

\item \textbf{VMN Name Similarity} - VMN is a string comparison function, which is used for matching two peoples' names (\cite{vosecky2009user}). It is designed for full and partial matches of names consisting of one or more words. VMN supports the case of swapped names and the cases of partial matches. We used VMN to calculate the similarity between the two users' full names.

\item \textbf{Names Frequency Similarity} - We used the frequency table of names in Facebook\footnote{\url{https://blog.skullsecurity.org/2010/return-of-the-facebook-snatchers}} to calculate the frequency of a particular name. We even decided to utilize the Facebook frequency list for user names from other social networks, since Facebook is the largest social network in the world with over a billion, monthly active users worldwide (\cite{fbstats2013}). Hence, there is a high probability that this table will also partially describe the frequency of names around the world. The similarity was calculated by the average number between the frequencies of the user name in each network. We decided not to normalize this score, and to let the machine learning model distinguish between high and low frequencies.   
\end{itemize}

\subsubsection{User Information Based Features}
User information-based features are features that represent the similarities between the different parts of the personal information of two users. In this study, we extracted the following 15 user information-based features for any two given profiles: 
\begin{itemize}
\item \textbf{Locations Distance} - In many social networks, such as Facebook, Twitter, and Google+ the location of the user is known. Since the location field is a complex characteristic, we cannot simply compare the names of the two locations. Therefore, we decided to calculate the distance between two locations, not to normalize it, and to let the machine learning model distinguish between short and long distances. We calculated this distance using Bing Maps.\footnote{\url{http://www.bing.com/maps/}} We calculated the distance between two hometowns and two current cities.

\item \textbf{N-Gram Current Employer Similarity} - We used the 3-grams method to calculate the similarity between the users' current employers.

\item \textbf{Damerau Levenshtein Current Employer Similarity} - 
We used this metric to calculate the similarity between the two users' current employers.

\item \textbf{Jaro-Winkler Current Employer Similarity} - We used this metric to calculate the similarity between the users' current employers.

\item \textbf{Jaccard Similarity } - The Jaccard measure  (\cite{jaccard1901etude}) simply measures the ratio of equal tokens in the union of tokens of both strings. The range of the Jaccard measure is between 0 and 1, 1 being an exact match and 0 otherwise. Misspellings are counted as different words and hence can decrease the score enormously. Moreover, Jaccard measure matches all kinds of words, even words that occur often, such as ``the'' and ``a.'' 
We used the Jaccard measure to create the following 4 features, by calculating the similarity between: 
\begin{itemize}
 \item \textbf{two users' full profiles without user names}.
 \item \textbf{two users' professional experiences}, if provided.
 \item \textbf{two users' educational backgrounds}, if provided.
 \item \textbf{two users' information page fields}. This was done in two steps. In the first step, for each profile in each network, we created a textual representation of the profile by aggregating all the user's information fields, which appeared in the user's profile page without the name, educational and background fields, to a single text. Next, in the second step, we compared the two texts using the Jaccard measure.
\end{itemize}
\item \textbf{Semi Vector Space Model (semi VSM) } - The Vector Space Model (VSM) (\cite{raghavan1986critical}) is a way of representing documents through the words that they contain and is a standard technique in Information Retrieval. The VSM allows decisions to be made regarding which documents are similar to each other for a given query. The vector space model procedure consists of three stages. The first stage is document indexing where content-bearing terms are extracted from the document text. The second stage is weighing the indexed terms in order to enhance the retrieval of documents that are relevant. The last stage ranks the document with respect to the query and according to a similarity measure. 
In order to use the VSM we needed sets of documents and a query.  In our case there are only two documents, and since our goal is to calculate the similarity between them, we used one user document as a query and the other user document as the only document retrieved. 
As we had only one retrieved document, we changed the implementation of each term weight by only calculating the TF value rather than multiplying the IDF value.
We used the semi VSM similarity to create the following 4 features, by calculating the similarity between:
\begin{itemize}
\item \textbf{two users' full  profiles without user names.}
 \item \textbf{two users' professional experiences}, if provided.
 \item \textbf{two users' educational backgrounds}, if provided.
\item \textbf{two users' information page fields} - this was done in two steps. In the first step, for each profile in each network, we created a textual representation of the profile by aggregating all the user's information fields, which appeared in the user's profile page without the name, educational and background fields, to a single text. Next, in the second step, we compared the two texts using semi VSM similarity.

\end{itemize}
Since we did not use the IDF value, we refer to this method as semi VSM.

\item \textbf{Vector Space Model of Full Profiles Similarity} - As described above, at first we did not use the IDF value. In order to use the VSM method as is, we decided that instead of calculating the similarity between two user documents only, we would calculate the similarity between one user document from the first social network and all the users' documents from the second network. To clarify the process, suppose $p$ and $\hat{p}$ are two users' documents where $p$ belongs to social network $S_1$ and $\hat{p}$ belongs to social network $S_2$. Instead of calculating the similarity only between $p$ and $\hat{p}$, we calculated the similarity between $p$ and all the users in $S_2$ and returned a score for $\hat{p}$; this is the similarity. The same applies to $\hat{p}$ for all the users in $S_1$. 
Notice that we did not refer to stop words in this category of features, since they are not frequent in our data.

\end{itemize}

\subsubsection{Social Network Topological Based Features}
\label{sec:top_based}
Social network topological based features are features that represent the similarity between the networks of two user profiles.
In this study, we extracted the following two social network based features for any two given profiles: 
\begin{itemize}
\item \textbf{Mutual Friends} - This feature represents the number of mutual friends (\cite{liben2007link}) between two users. Mutual friends are counted by the number of friends with identical names in both circles of friends. 
\item \textbf{Mutual Friends of Friends} - This feature represents the number of mutual friends of friends of two users. In this study, we defined mutual friends to be the friends with identical names in both cycles of friends. 
\end{itemize}

\subsection{Constructing Training and Testing Datasets}
\label{sec:constructing_training}
To construct all the various training and testing datasets, we performed the following steps: First, we manually matched pairs of users which were members of both networks. Secondly, we used the matched pairs of users to construct training and testing datasets. Lastly, for each pair of users in each of the training and testing datasets, we extracted two different sets of features, vector based on the features describe in Section~\ref{sec:features}. In the following sections, we describe each of these steps in detail.

\subsubsection{Matching Pairs of Users}
To identify which users were members of both networks, we performed a cross-reference between the collected user's profiles according to their names. We then used the profile photos, and manually checked each matched pair of profiles to see whether they belonged to the same real person. Each checked pair was labeled as a match or a non-match. The remaining users who did not have a match by name were used to create negative pairs. 

\subsubsection{Constructing Datasets}
After we created pairs of matched and unmatched users, the next step was to create training and testing datasets. 

We constructed the training and testing datasets by using the following method: First, to build the training dataset we randomly divided all the users we crawled from both networks into ten groups and associated each labeled pair with the appropriate group (i.e., the group that contains both of the users of the labeled pair). Note that if the users of a labeled pair belonged to a different group, the labeled pair was thrown out. Next, for each positive pair in each group, we randomly generated ten negative pairs, namely, where $<p_1, \hat{p}_1>$ is a matched pair, with $p_1$ belonging to social network $S_1$ and $\hat{p}_1$ belonging to social network $S_2$. Our main assumption is that a real person has only one user profile page per OSN site. Therefore, there is no user in $S_1$ who can also be matched to $\hat{p}_1$ and there is no user in $S_2$ who can also be matched to $p_1$. Using this assumption, for $p_1$, we randomly picked five user profiles from $S_2$ that are different from $\hat{p}_1$; and for $\hat{p}_1$, we randomly picked five user profiles from $S_1$ that are different from $p_1$. These pairs obviously did not match, and were therefore labeled as negative pairs. Then, in order to build ten training datasets and ten testing datasets, we took out nine different groups from the ten groups and used them as a training dataset, and the group that was left was used as a testing dataset. Finally, for each pair of users in each of the constructed training and testing datasets, we calculated all the features described in Section~\ref{sec:features}.
It is worth mentioning that in this study,
in contrast to related studies (\cite{fire2013computationally,fire2013friend}), we could not use the cross-validation approach (\cite{kohavi1995study}) as is because we needed to ensure that there would not be any overlap of users in each phase of the cross-validation. If we were to use the classical 
cross-validation process, all the pairs would be split into training and testing datasets in each one of the iterations. 
For example, suppose that the pair of profiles $<p_{1}, \hat{p}_{1}>$ is labeled as a positive match, and the pair of profiles $<p_{1},\hat{p}_{2} >$ is labeled as negative match. Also suppose that in one iteration of cross validation $<p_{1}, \hat{p}_{1}>$ were in a training dataset and $<p_{1},\hat{p}_{2} >$ were in a testing dataset. Here there is a dependency between the training and testing datasets that is inconsistent with the cross validation method. Therefore, to overcome the issue of users overlapping, we decided to manually construct ten training datasets and ten testing datasets.

\subsection{Building the Model}
In this step the goal is to build a model, which can calculate, for two user profiles from different OSNs, the probability that they belong to the same real individual. To build such a model we used machine learning techniques. In the real world there is a large ratio of people who have several accounts on different social networking sites compared to those who do not. To simulate this ratio, we used a ratio of 1 to 5; meaning that for every positive pair there were 5 negative pairs (see Section~\ref{sec:constructing_training}). 
One of the main difficulties in building a model is choosing the right algorithms to apply to the training dataset. In our case, the data was overwhelmingly imbalanced and therefore, we needed to use algorithms that had the capability of dealing with imbalanced data.  We chose to use six different algorithms, which were better suited to deal with imbalanced data, namely, AdaBoost (\cite{freund1996experiments}), Rotation Forest (\cite{rodriguez2006rotation}), Random Forest (\cite{breiman2001random}), Logistics Model Tree (LMT)  (\cite{landwehr2005logistic}), LogitBoost (\cite{Friedman98additivelogistic}), and Artificial Neural Networks (\cite{yao1999evolving}). We constructed our classifiers using Weka (\cite{weka}), a popular suite of machine learning,  and the features defined in Section~\ref{sec:features}.
  
For each of these algorithms, most of the configurable parameters were set to their default values. We bounded the process in AdaBoost by 200 iterations and used a classification tree as the base inducer. For Random Forest we employed an ensemble of 150 random classification trees. Also we used the LMT with a minimum number of instances of 30 and the LogitBoost with 25 iterations.


\subsection{Entity Matching Scenarios}
\label{sec:scenarios}
In this study, we utilized the above described extracted features, training and testing datasets, and the six machine learning algorithms to perform entity matching between two profiles for the following three scenarios: 
(a) \textbf{Matching users across two social networks } - this scenario is the most general one. Given two users' profiles from different social
networks, we calculated the probability that these two users' profiles belonged to the same real individual using all the features defined in Section~\ref{sec:features}; 
(b) \textbf{Searching for a user} - in this scenario we simulated a case in which we had a person's name, and  we wanted to find this person's accounts containing a similar name, on different social networks. To simulate this scenario, we removed all pairs of users that do not have similar names from the testing set. In other words,
the test set contained only pairs of users who had similar names (Soundex name similarity is equal to one); and
(c) \textbf{De-anonymizing user's identity} - 
in this scenario we simulated a case in which we had a person's name and profile details on one social network, and  we wanted to find his or her hidden accounts which contain a pseudonym, on different social networks.
To simulate this scenario, we used all features defined in Section~\ref{sec:features}, except the named based features.


\subsection{Evaluation Measures}
To compare the various constructed classifiers' performances in each of the above scenarios, we used the accuracy, true positive rate (TPR) and false positive rate (FPR), AUC measures. The accuracy measure represents the percentage of test instances correctly classified. It has some disadvantages as a performance estimate, namely, it is not chance-corrected and it is not sensitive to class distribution. Therefore, in our case it is a less significant measure. The true positive rate represents the proportion of positive cases correctly identified. The false positive rate represents the proportion of negative cases that were incorrectly classified as positive. The  Receiver Operating Characteristic (ROC) is a standard technique for summarizing classifier performances over a range of trade-offs between true positive rates, and false positive error rates (\cite{fawcett2006introduction}). Each point in the curve corresponds to a particular cut-off, where the x-value is the false positive value (1-specificity), and the y-value is the sensitivity value. Points closer to the upper right hand corner correspond to lower cut-offs and points closer to the lower left hand corner correspond to higher cut-offs. The choice of the cut-off thus represents a tradeoff between sensitivity and specificity. In terms of classifier comparison, the best curve is the leftmost one, coinciding with the y-axis. Therefore, the AUC is an accepted performance metric for a ROC curve (\cite{fawcett2006introduction}). The AUC range is $[0, 1]$. The area under the diagonal is 0.5 and this value represents a random classifier. A value of 1 represents an optimal classifier. The main advantage of the AUC measure is that unlike other accuracy measures, the AUC is not influenced by the imbalance distribution of the classes (\cite{menon2011link}).

\subsection{Features Analysis}
\label{sec:feature_analysis}
To evaluate the contribution of each feature in the first entity matching scenario, we performed two
procedures for each feature: ``all-but-x'' and ``only-x.'' The first procedure ``all-but-x'' aims to
measure how much a certain feature contributes to the entire model. 
However, the second procedure, ``only-x,'' aims to evaluate how each feature performs on its own. 
To calculate each feature's merit using the ``all-but-x'' procedure, we performed the following steps:
First, we selected the classification algorithm that obtained the highest AUC measure in the matching users across two social networks scenario (see Section~\ref{sec:matching_results}). Secondly, using the selected algorithm, we constructed classifiers using all the features except for one and measured the decline in AUC. Lastly, we used the obtained AUCs as measures for comparing the performance of various features.

To calculate each feature's merit using the ``only-x'' procedure, we performed the following steps:
First, we selected the classification algorithm that obtained the highest AUC measure in the matching users across two social networks scenario (see Section~\ref{sec:matching_results}). Secondly, using the selected algorithm, for each feature, we constructed classifiers using only that feature. Then, we measured the constructed classifier's AUC. Lastly, we used the obtained AUCs as measures for comparing the performance of various features.

\section{Experimental Study}
\label{sec:exper}
\subsection{Dataset}
\label{sec:dataset}
To retrieve real data we chose to acquire data from two large and popular social networks, Facebook and Xing, by using dedicated web crawlers. Facebook is a free, popular social networking website that was launched on February 2004. As of December 2013, Facebook has 1.23 billion monthly active users (\cite{fbstats2013}). Facebook allows registered users to create profiles, upload photos and videos, send messages, and keep in touch with friends, family, and colleagues.
Xing, a European social network for business professionals, founded in 2003 in Hamburg, Germany, had reached 13.76 million users as of September 2013 (\cite{xingstat2013}). Xing offers personal profiles, contacts, groups, discussion forums, event coordination, and other common social community features. Xing is a platform where professionals from all kinds of different industries can meet, find jobs, colleagues, new assignments, cooperation partners, experts, and generate business ideas. Xing is particularly popular in Switzerland, Germany, and Austria (\cite{bilge2009all}). Additionally, Xing competes with the American platform LinkedIn for social networking among businesses. We developed a dedicated crawler component to crawl through Facebook and Xing and collect information on user profiles and contact lists accessible to the public. The crawler starts with a user's profile page, downloads it, downloads its friends list, and then continues to crawl each friend, friend of friend, and so on.
To gain access to public profiles on each of the social networks, we had to use a real account on each network. To choose the starting profiles of each crawler, we manually found pairs of public profiles, one from each network, that belonged to the same real individual. Using these starting profiles ensured that we would have some mutual users across these two social networks. For each user profile we stored the user profile page and the page of the user's friends, which both are formatted in HTML. In order to use the raw data that we crawled, the data required  preprocessing. In our case, based on the knowledge of the structure of the social network, content was extracted and irrelevant data was filtered out. We developed a tool that extracted interesting personal data about each user of each network, such as name, gender, professional experience, education, a list of friends' names, etc. All the data that was extracted from each network is described in Table~\ref{tab:details}.
\begin{table}

    \caption{\textbf{Extracted data from Facebook and Xing}}

	\begin{center}
  \begin{tabular}{c}

      \includegraphics[scale=0.5]{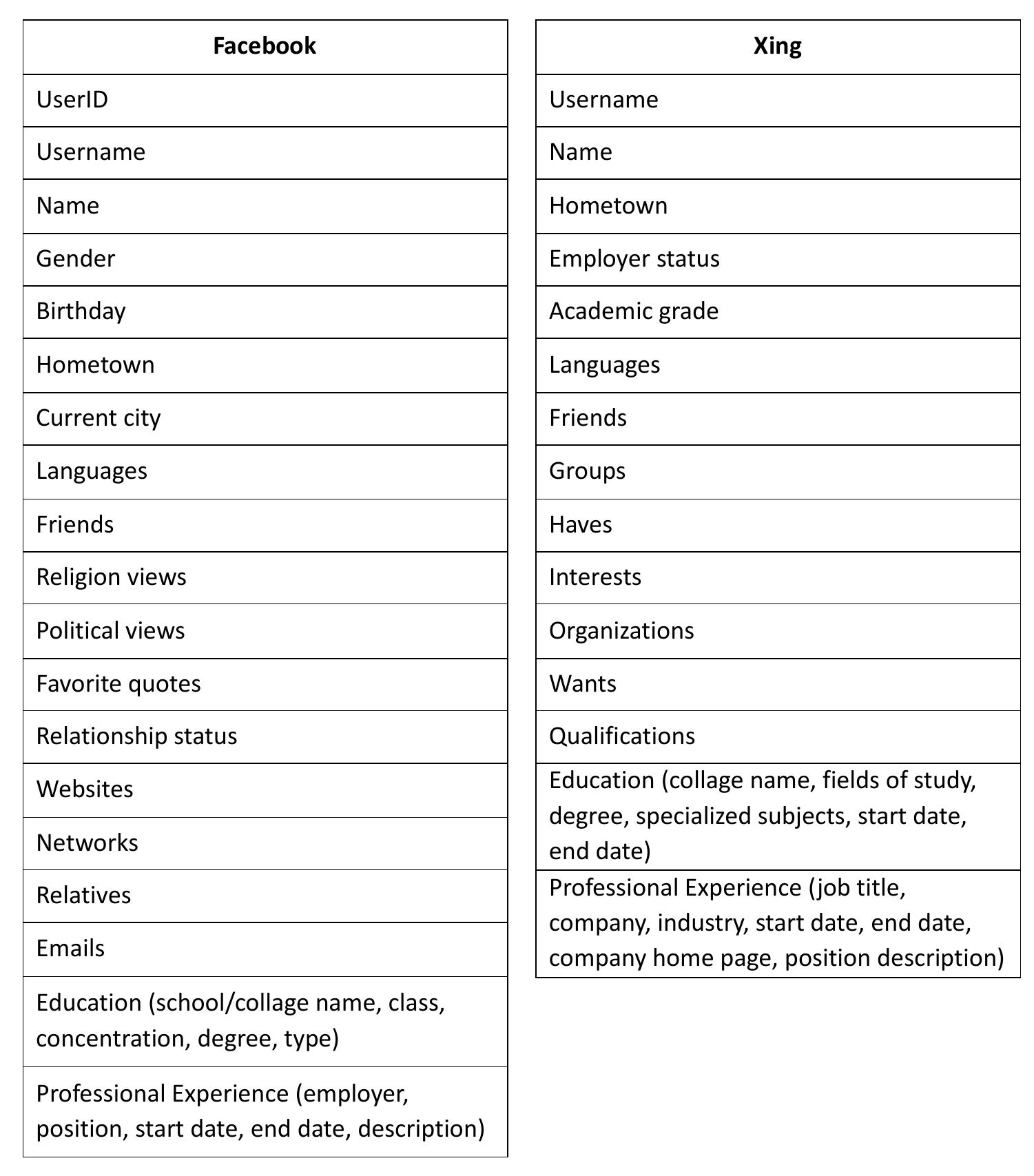}

  \end{tabular}
	\end{center}
	    \label{tab:details}

\end{table}

After the preprocessing step we decided to store the data in a relational data model using the database. As we were dealing with real data from people, we were able to mark which profiles from the two social networks belonged to the same real person. This point was particularly  important because we were able to compare the outcome of the matching process with this ground truth. 
We collected 16,561 user profiles from Facebook and 15,430 from Xing. Among the collected users, there were 464 pairs of users, each pair consists of two users one Xing user and one Facebook user, where the Xing user's full name contains the Facebook user's full name (we will refer to these users as users with very similar names). Each of the 464 pairs were manually checked to discover whether they represented two user profiles that belonged to the same real entity. Finally, we received 158 pairs of users with very similar names who belonged to the same real entity and 306 pairs of users with very similar names that did not belong to the same real entity. 
Using the crawled data, we constructed the training and testing datasets using the methods described in Section~\ref{sec:constructing_training}. 

\subsection{Results}
\label{sec:results}
In the following subsections, we present the results obtained utilizing the algorithms and methods described in Section~\ref{sec:method} on the constructed Facebook and Xing training and testing datasets. 
The results consist of two parts:  First, in Section~\ref{sec:eval_results}, we present the obtained results of the  entity matching classifiers on each one of the three entity matching scenarios that we described in Section~\ref{sec:scenarios}.  Secondly, in Section~\ref{sec:features_results}, we present the results of the feature analysis we performed according to procedures we described in  Section~\ref{sec:feature_analysis}. 

\subsection{Scenarios Evaluation Results}
\label{sec:eval_results}
As described in Section~\ref{sec:method}, we  used various machine learning algorithms to construct entity matching classifiers, which can match entities across two social networks. In the following three subsections, we present the constructed classifiers' results for each one of the entity matching scenarios which we described in Section~\ref{sec:scenarios}.

\subsubsection{Matching Users Across Two Social Networks}
\label{sec:matching_results}
\begin{figure}[ht]

\centering
\begin{subfigure}[AUC results]{\includegraphics[width=.45\linewidth] {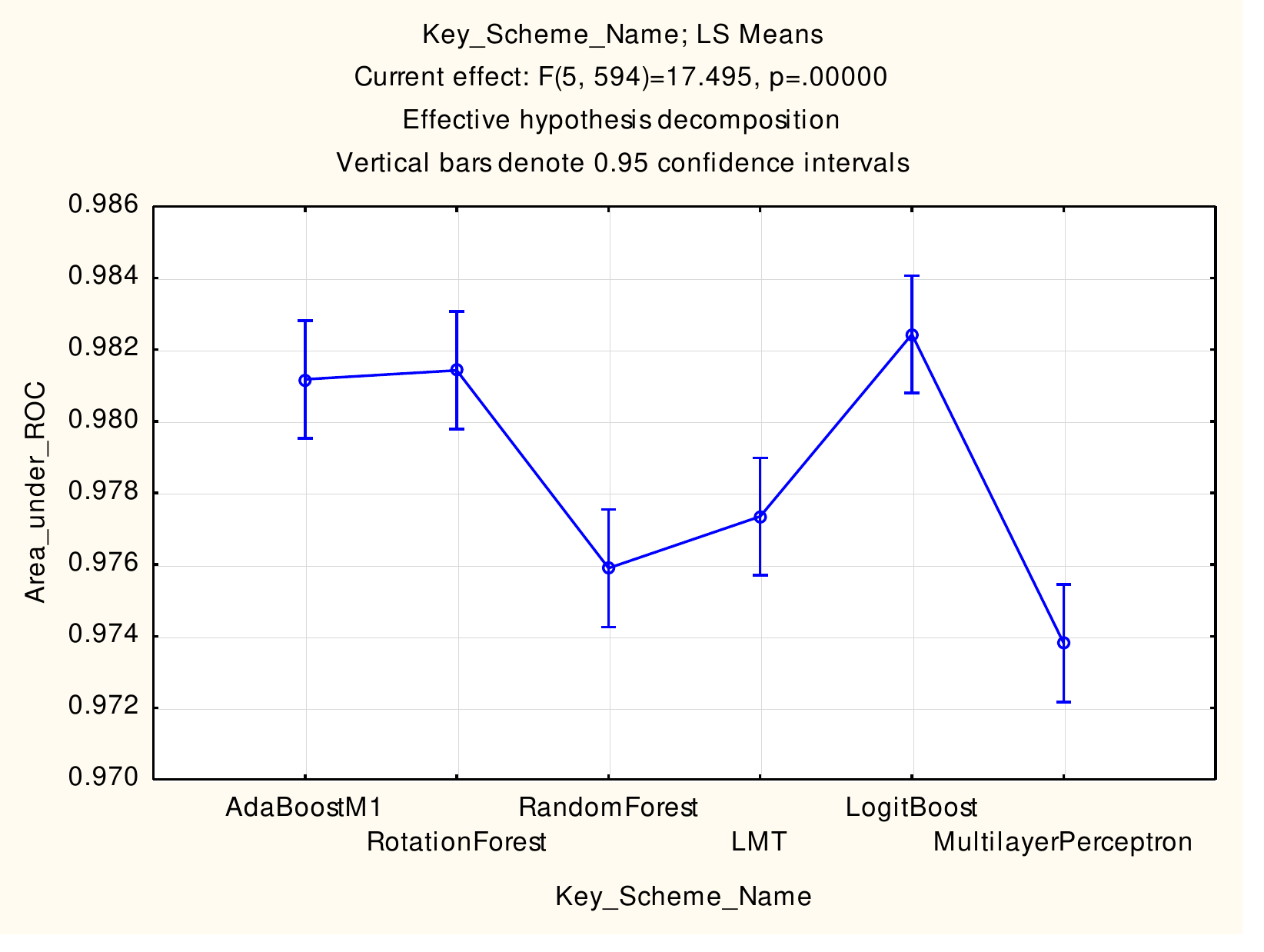}
   \label{fig:all_auc}
 }%
\end{subfigure}\hfill
 \begin{subfigure}[Accuracy results]{\includegraphics[width=.45\linewidth]{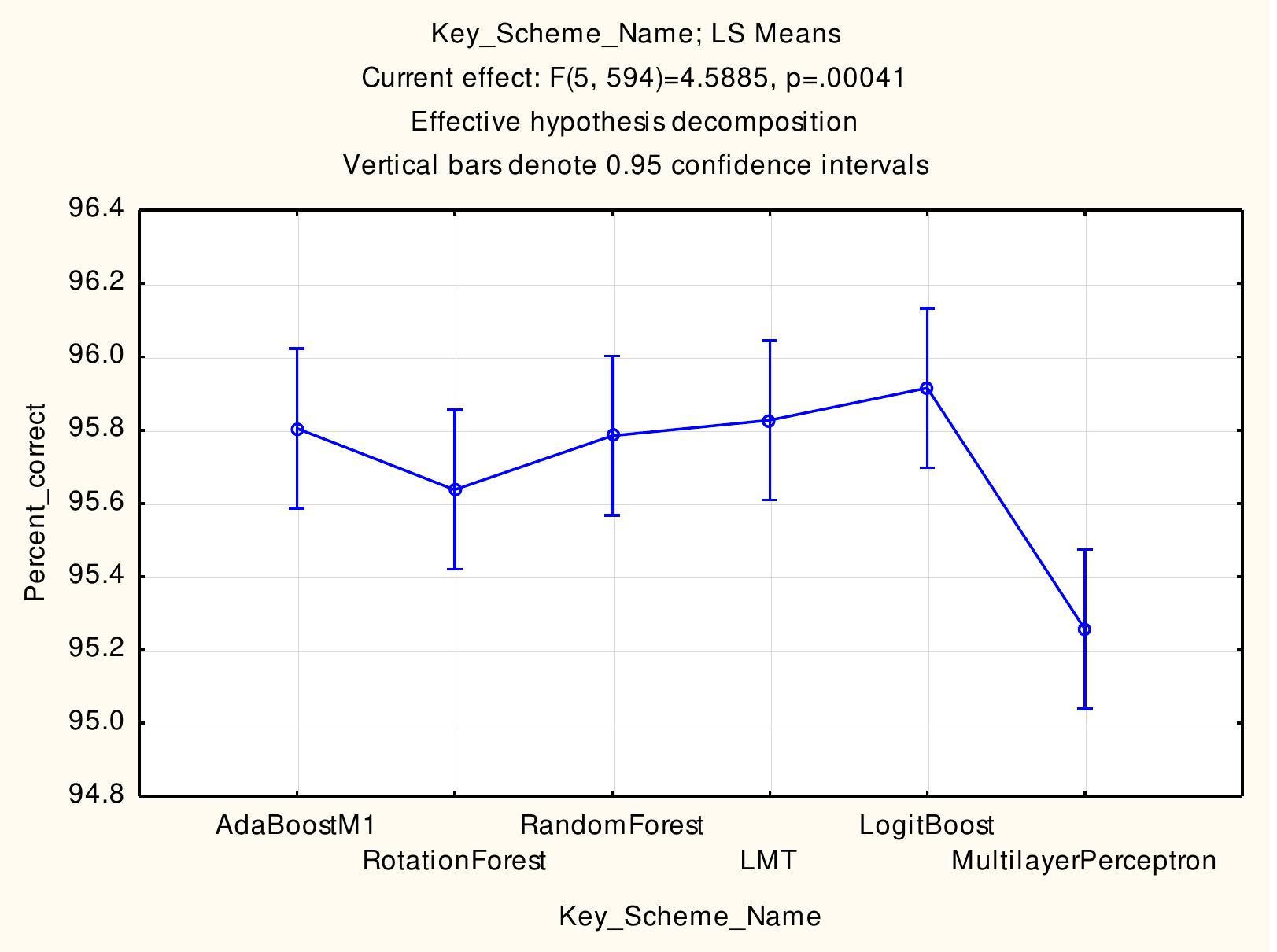}
   \label{fig:all_accuaracy}
 }%
\end{subfigure}\\
\begin{subfigure}[True positive rate results]{\includegraphics[width=.45\linewidth] {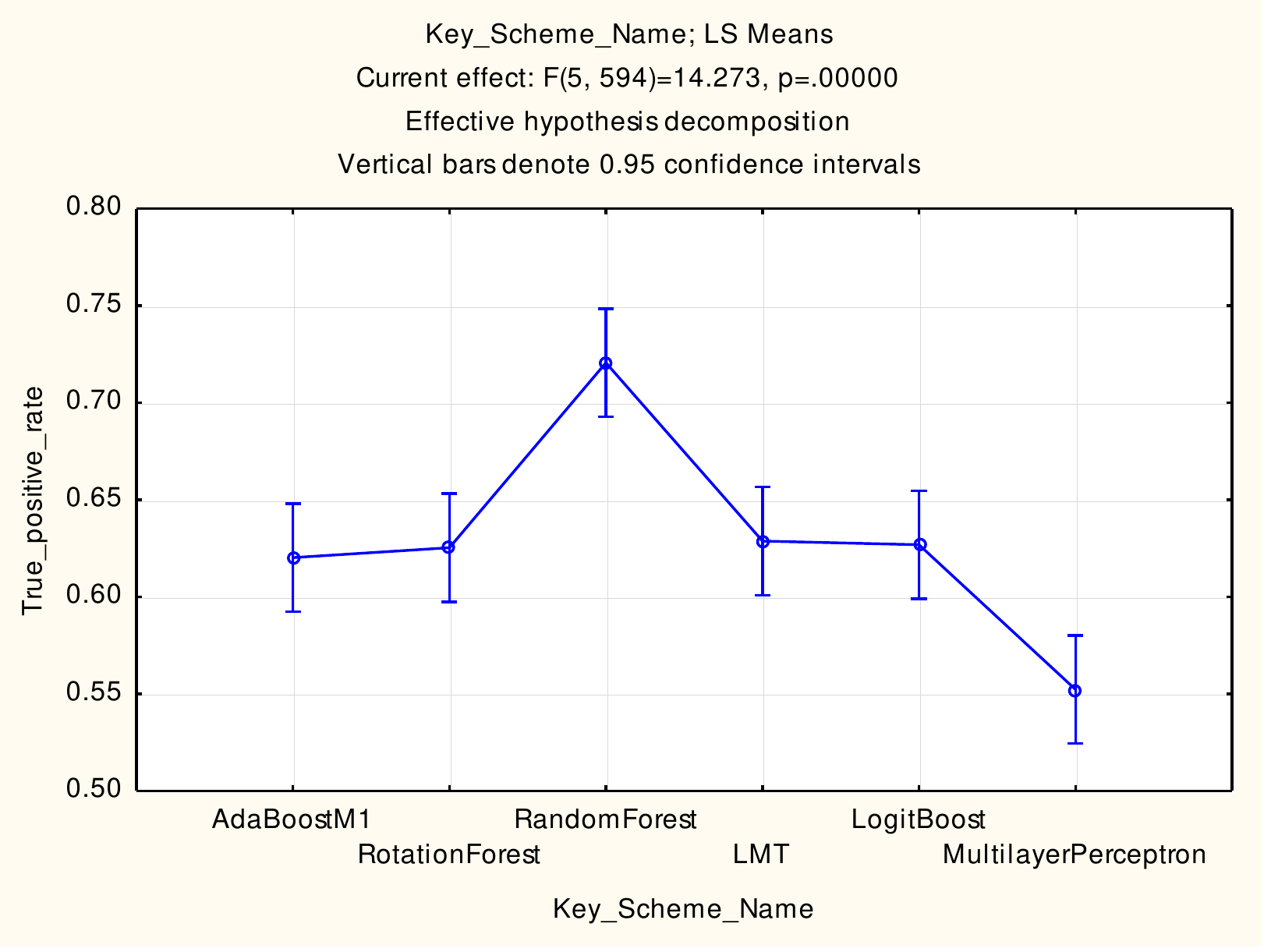}
   \label{fig:all_tpr}
 }%
\end{subfigure}\hfill
 \begin{subfigure}[False positive rate results]{\includegraphics[width=.45\linewidth]{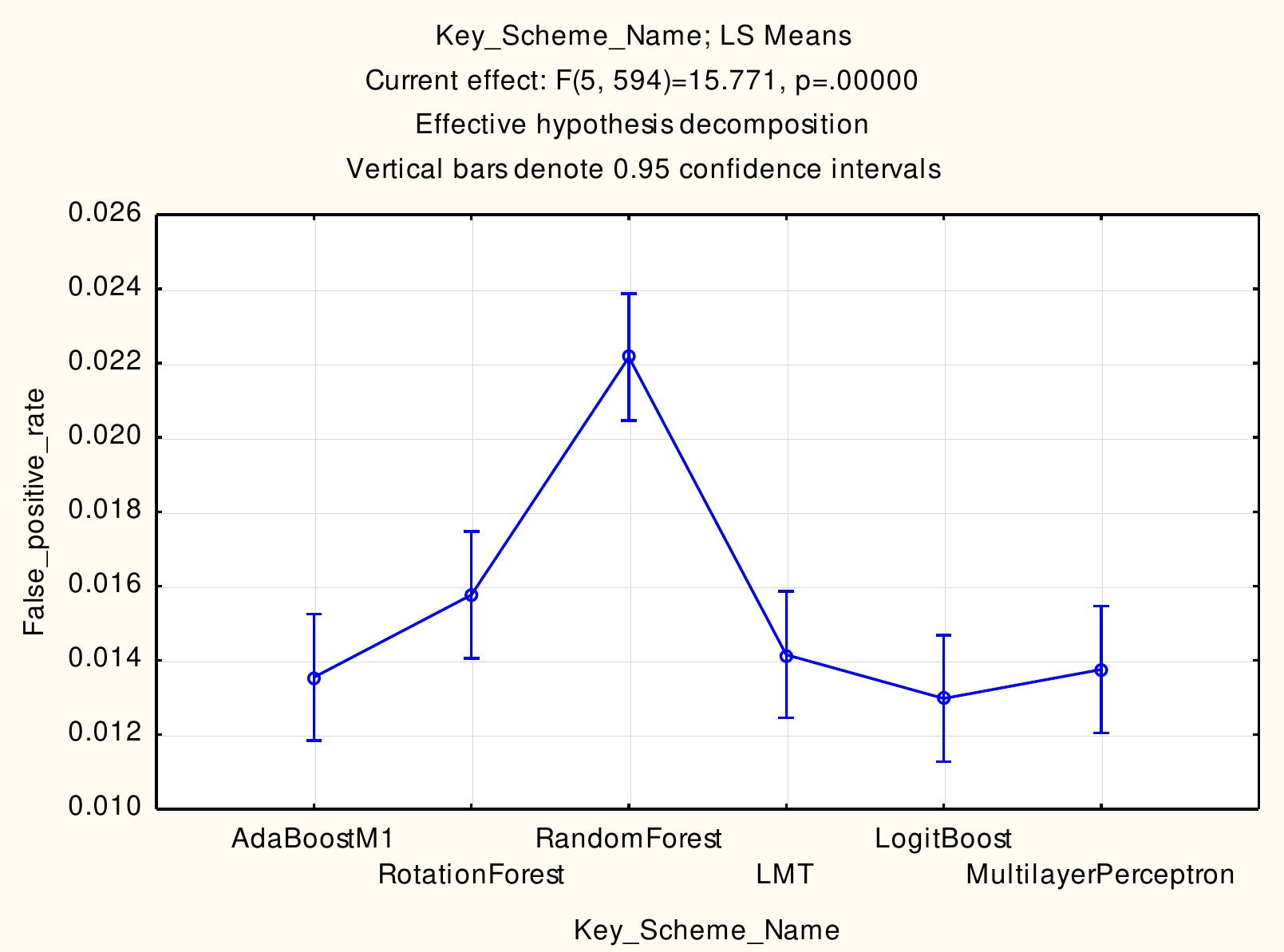}
   \label{fig:all_fpr}
 }%
\end{subfigure}

\caption{Matching users across two social networks using all features results}
\label{fig:all_results}
\end{figure}

The six classifiers' performance results in terms of AUC, accuracy, TPR, and FPR in this scenario are presented in Figure~\ref{fig:all_results}.
As shown in Figure~\ref{fig:all_auc}, all of the algorithms' AUCs were above $0.972$, where the LogitBoost classifier demonstrated the best performance and significantly outperformed the Multilayer Perceptron, LMT, and Random Forest classifiers. 
The null-hypothesis that all methods' performances were equal and the observed differences among their AUC values were merely random, was rejected using the ANOVA test, with $F(5,594)=17.495, p<0.01$. Even though there are small differences between the six methods' AUC, the differences are still statistically significant.
Figure~\ref{fig:all_accuaracy} presents the classifiers' accuracy results, which are consistent with the reported AUC results. Namely all methods reached a high accuracy of above 95\% but LogitBoost outperforms all others methods (according to the ANOVA test with $F(5,594)=4.5885, p<0.01$). 

Unlike the AUC and the accuracy results, the Random Forest classifier demonstrated the best TPR performance and significantly outperformed all other classifiers with a TPR of 0.72 (see Figure~\ref{fig:all_tpr}).  This means that 0.72 of the matched pairs were correctly identified.  The null-hypothesis, namely, that all methods' performances were equal and the observed differences among their true positive rate values are merely random, was rejected using the ANOVA test, with $F(5,594)=14.273, p<0.01$.
Additionally, as it shown in Figure~\ref{fig:all_fpr}, all of the classifiers' FPRs are under 0.024. This denotes that only 0.024 of  unmatched pairs were incorrectly identified. Furthermore, these results support the AUC and accuracy results, which show that the LogitBoost classifier demonstrated the best performance and significantly outperformed the Random Forest classifier.  The null-hypothesis, namely, that all methods' performances were equal and the observed differences among their FPR values are merely random, was rejected using the ANOVA test, with $F(5,594) =15.771, p<0.01$. These results indicate that although LogitBoost classifier's TPR was not as high as RandomForest classifier's TPR, it presented a much better FPR.	

\subsubsection{Searching for a User}
\label{sec:searching_results}
\begin{figure}[ht]

\centering
\begin{subfigure}[AUC results]{\includegraphics[width=.45\linewidth] {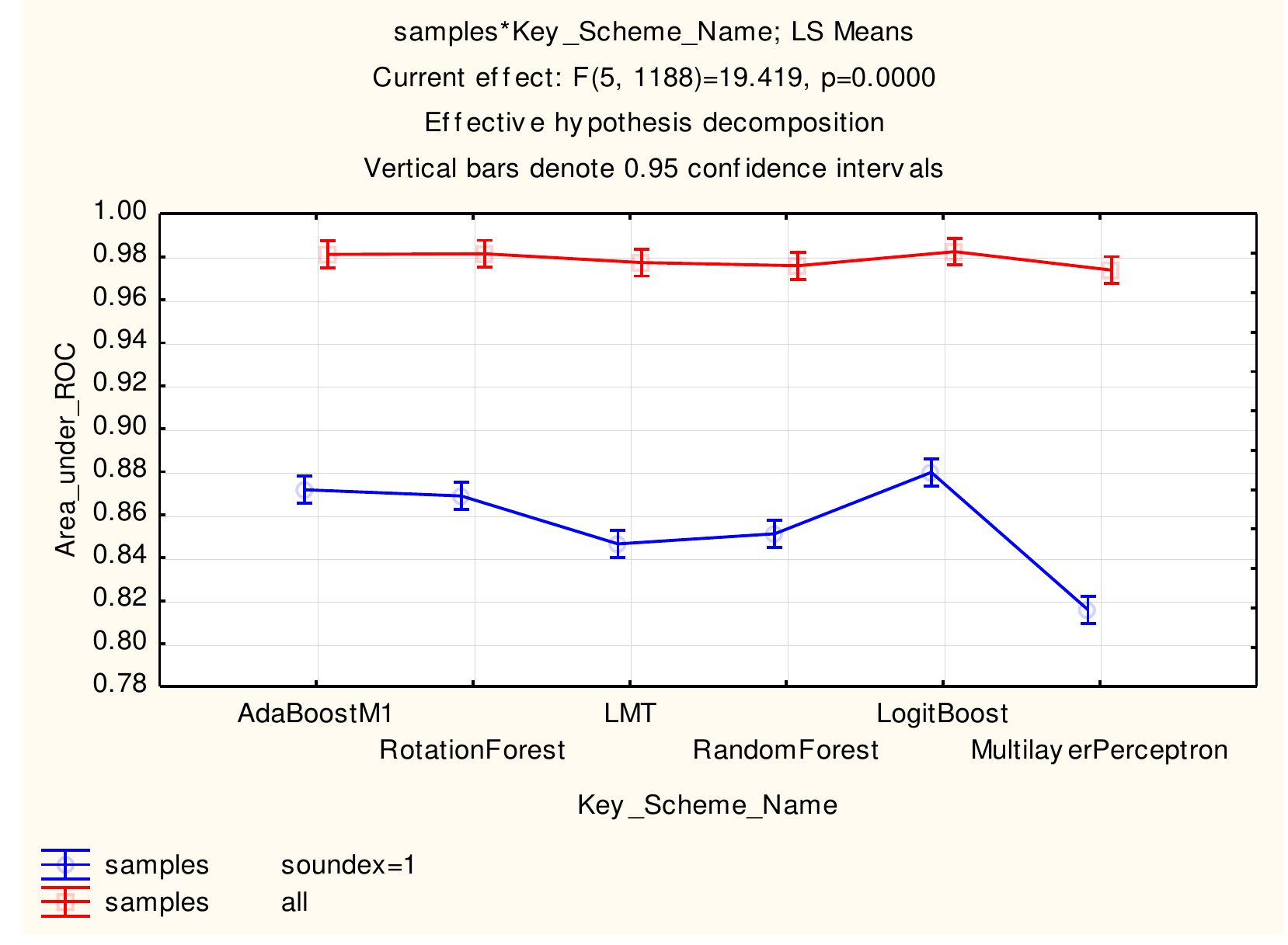}
   \label{fig:search_auc}
 }%
\end{subfigure}\hfill
 \begin{subfigure}[Accuracy results]{\includegraphics[width=.45\linewidth]{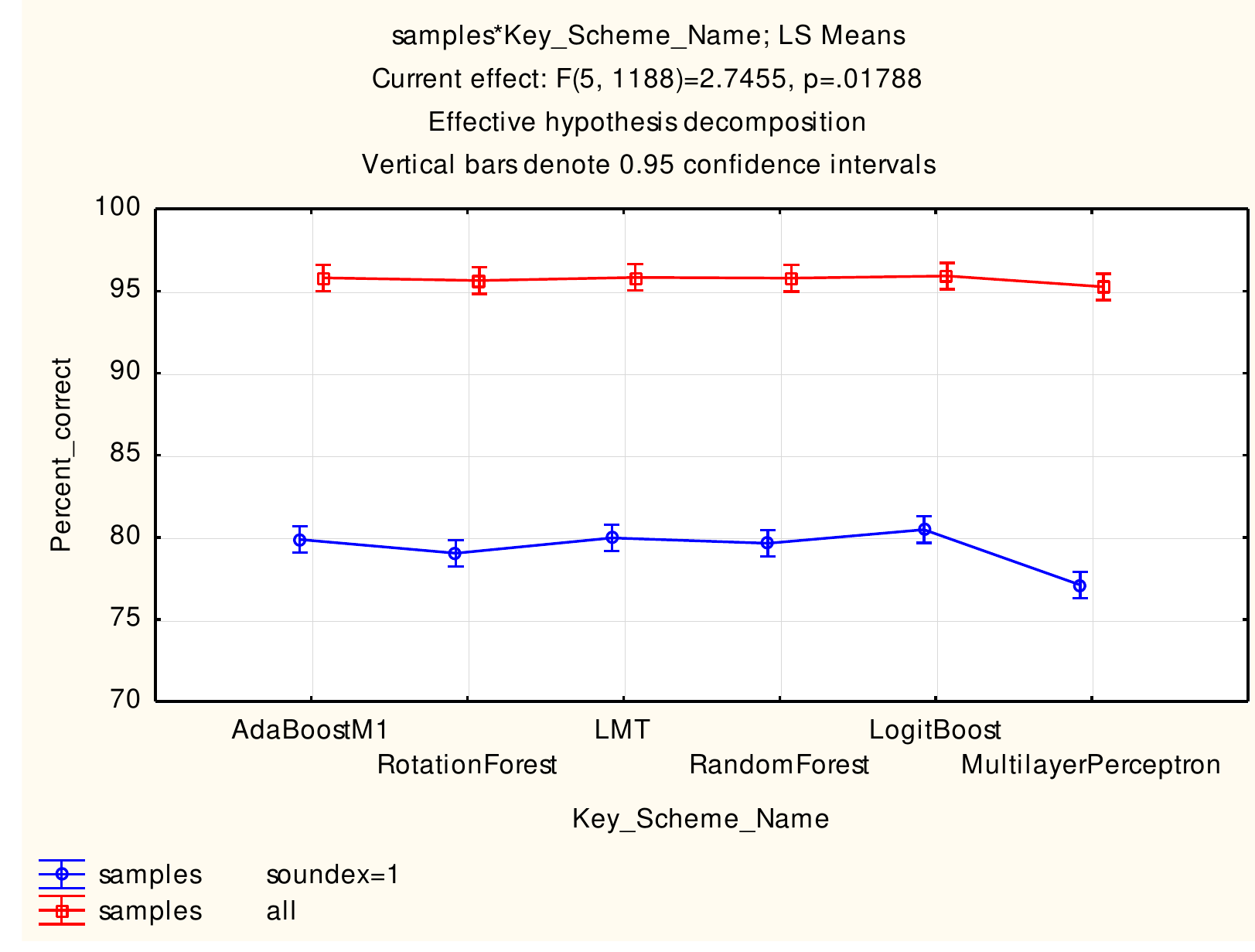}
   \label{fig:search_accuaracy}
 }%
\end{subfigure}\\
\begin{subfigure}[True positive rates results]{\includegraphics[width=.45\linewidth] {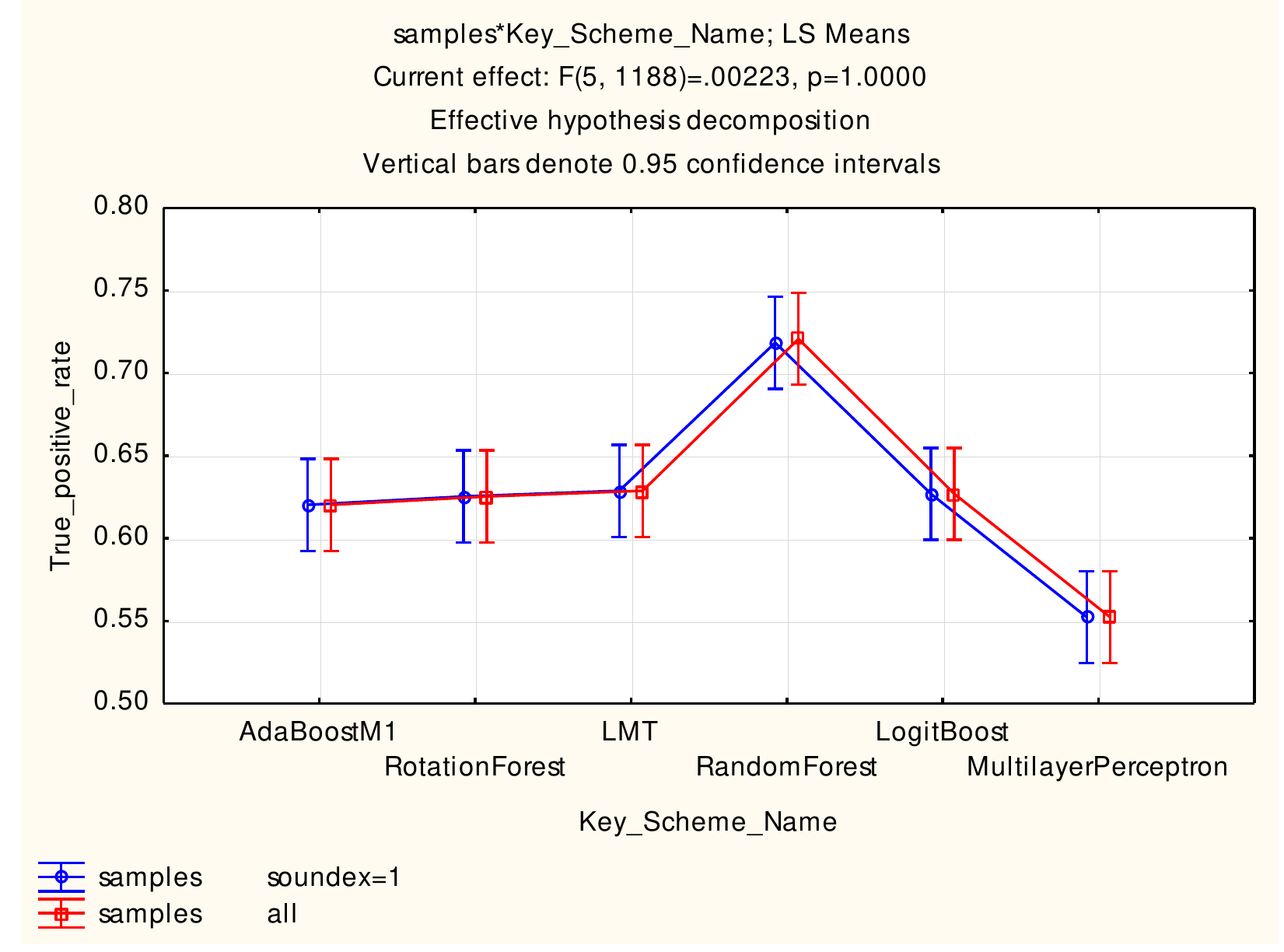}
   \label{fig:search_tpr}
 }%
\end{subfigure}\hfill
 \begin{subfigure}[False positive rates results]{\includegraphics[width=.45\linewidth]{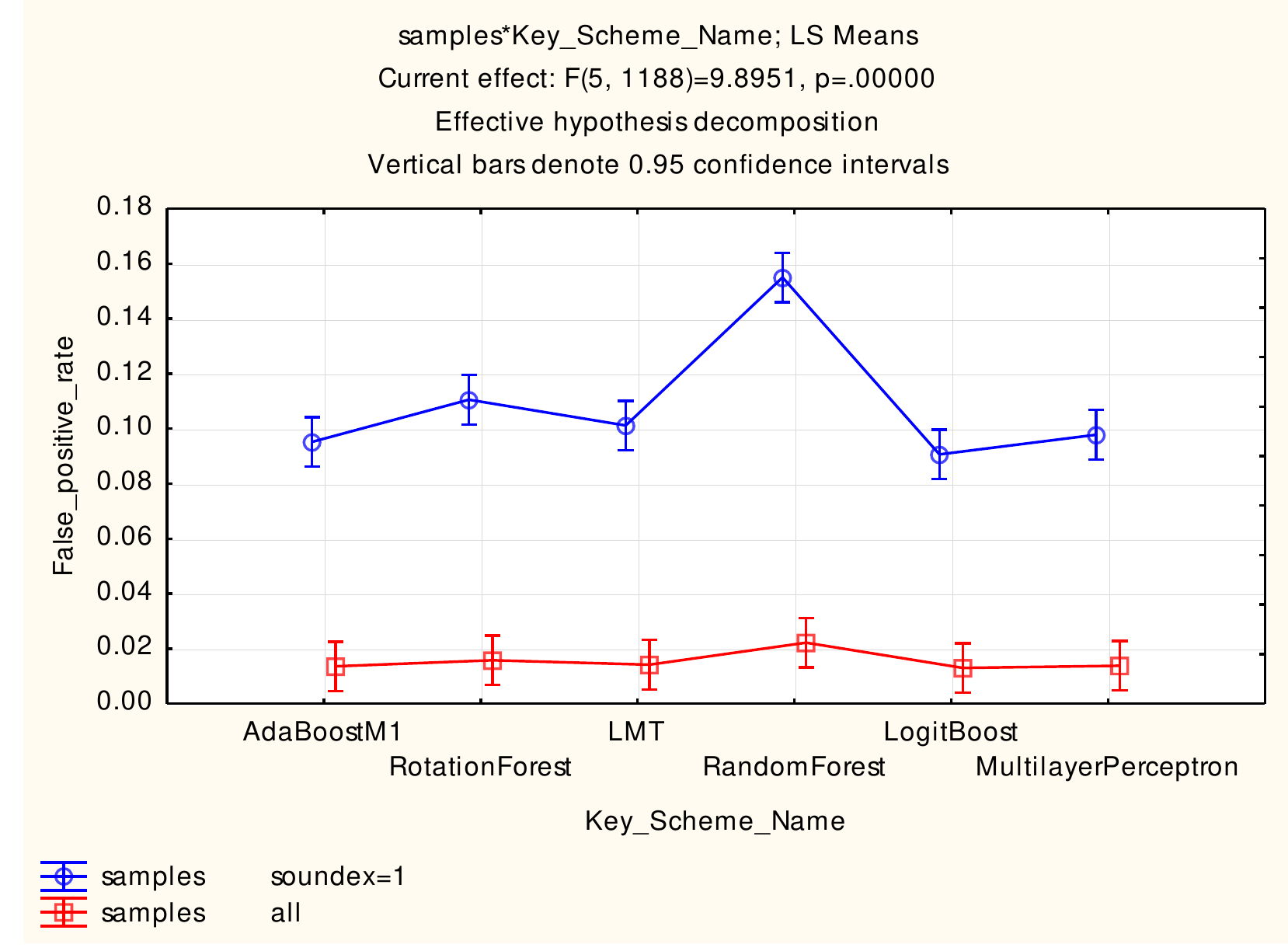}
   \label{fig:search_fpr}
 }%
\end{subfigure}

\caption{Searching for a user scenario results (marked in blue) compared to matching users across two social networks scenario results (marked in red)}
\label{fig:search_results1}
\end{figure}

In this scenario we supposed that we knew the user's full name and we wanted to find the user's accounts on different social networks. To simulate this scenario, we removed all pairs of users that did not have similar names from the testing dataset. In other words, the testing dataset consisted only of pairs of users who had similar names, with the Soundex name similarity equal to one.
The six classifiers' performance results in terms of AUC, accuracy, TPR, and FPR in this scenario are presented in Figure~\ref{fig:search_results1}.

As is displayed in Figure~\ref{fig:search_auc}, in this scenario all the classifiers' AUCs (marked in blue) were above $0.8$, where the LogitBoost classifier demonstrates the best performance and significantly outperforms the  Multilayer Perceptron, LMT, and Random Forest classifiers.  The null-hypothesis, namely, that all classifiers' performances were equal and the observed differences among their AUC values are merely random, was rejected using the ANOVA test, with $F (5, 1188) =19.419, p<0.01$. Additionally, we also compared the classifiers' accuracy results (see Figure~\ref{fig:search_accuaracy}), and the classifiers' accuracy results were consistent with the reported AUC results. Namely all classifiers reached an accuracy of above 75\% but the LogitBoost classifier outperformed all other classifiers (according to the ANOVA test with $F (5, 1188) =2.7455, p<0.01$).

Figure~\ref{fig:search_tpr} illustrates the classifiers' TPR results (marked in blue) in this scenario. Unlike the AUC and the accuracy results, the Random Forest classifier demonstrated the best performance and significantly outperformed all the methods.  The null-hypothesis, namely, that all methods' performances were equal and the observed differences among their true positive rate values are merely random, was rejected using the ANOVA test, with $F(5, 1188) =0.00223, p<0.01$. Additionally, Figure~\ref{fig:search_fpr} presents the classifiers' FPR results in this scenario. As depicted, all of the classifiers' FPRs were under 0.16.  This supports the AUC, and the accuracy results show that the LoogitBoost classifier demonstrated the best performance and significantly outperformed the Rotation Forest and Random Forest classifiers.  The null-hypothesis, namely, that all methods' performances were equal and the observed differences among their false positive rate values are merely random, was rejected using the ANOVA test, with $F(5, 1188) =9.8951, p<0.01$. 

\subsubsection{De-anonymizing User's Identity}
\label{sec:de_results}
\begin{figure}[ht]
\centering
\begin{subfigure}[AUC Results]{\includegraphics[width=.45\linewidth] {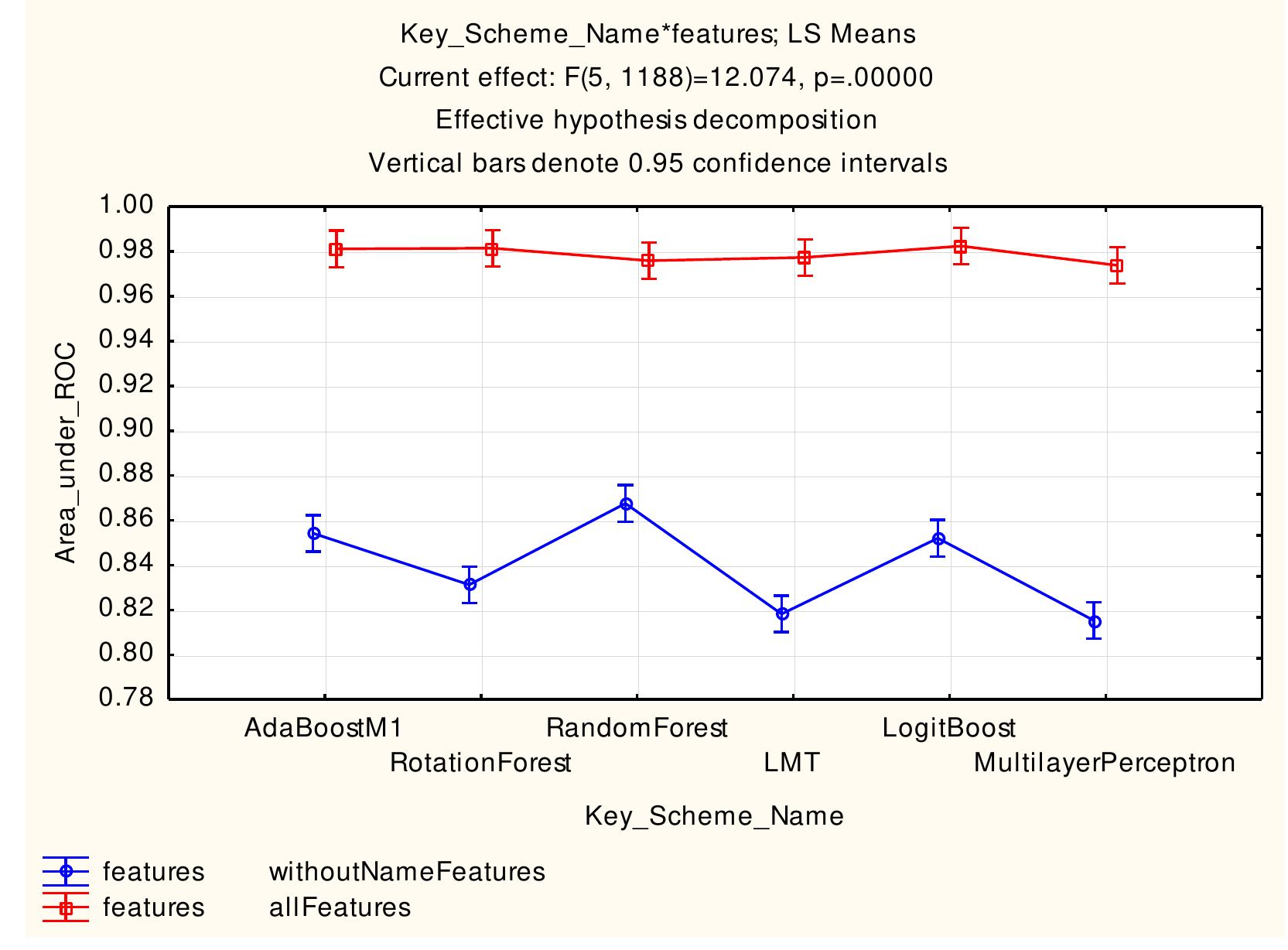}
   \label{fig:de_auc}
 }%
\end{subfigure}\hfill
 \begin{subfigure}[Accuracy Results]{\includegraphics[width=.45\linewidth]{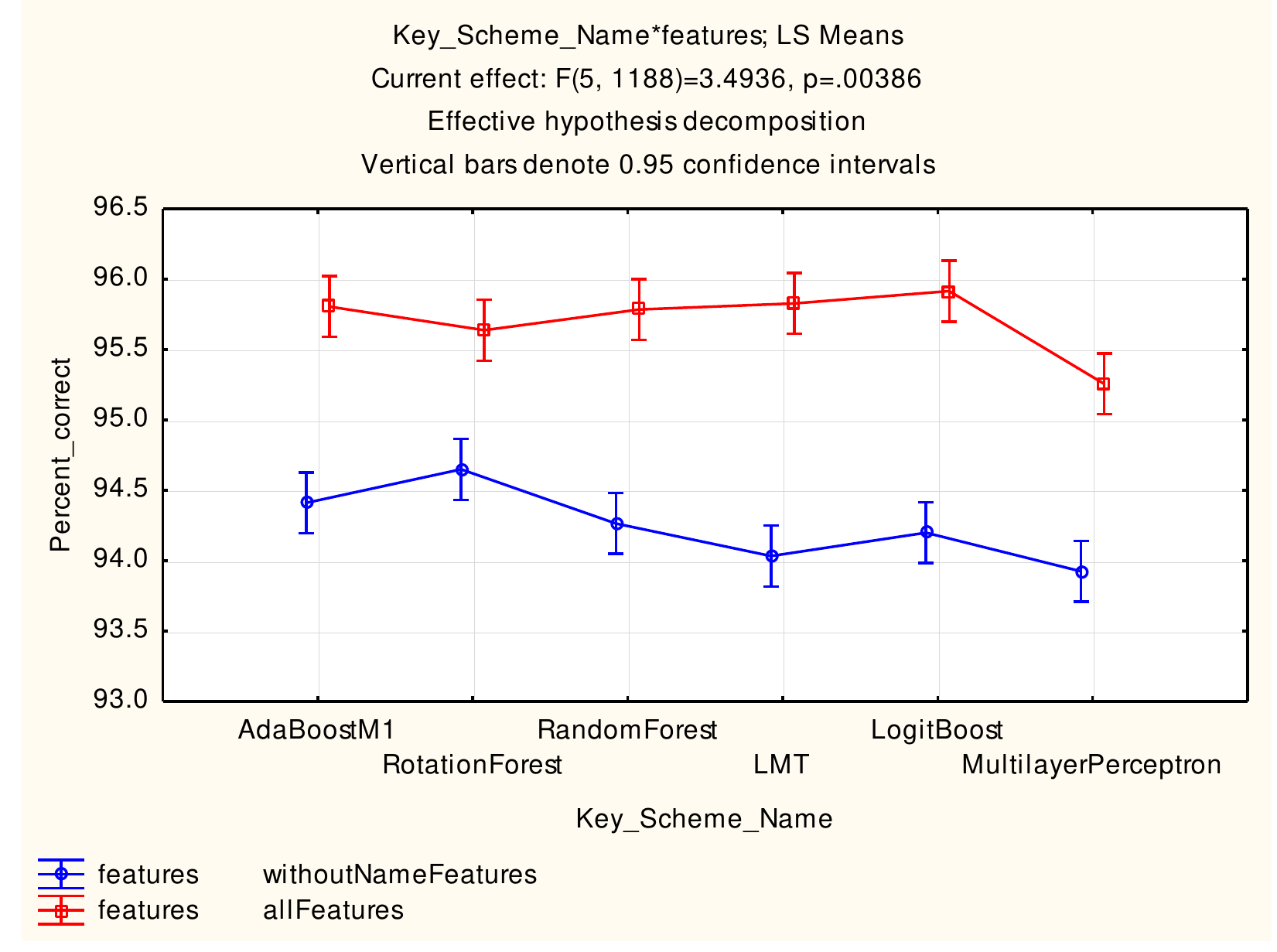}
   \label{fig:de_accuaracy}
 }%
\end{subfigure}\\
\begin{subfigure}[True Positive Rates Results]{\includegraphics[width=.45\linewidth] {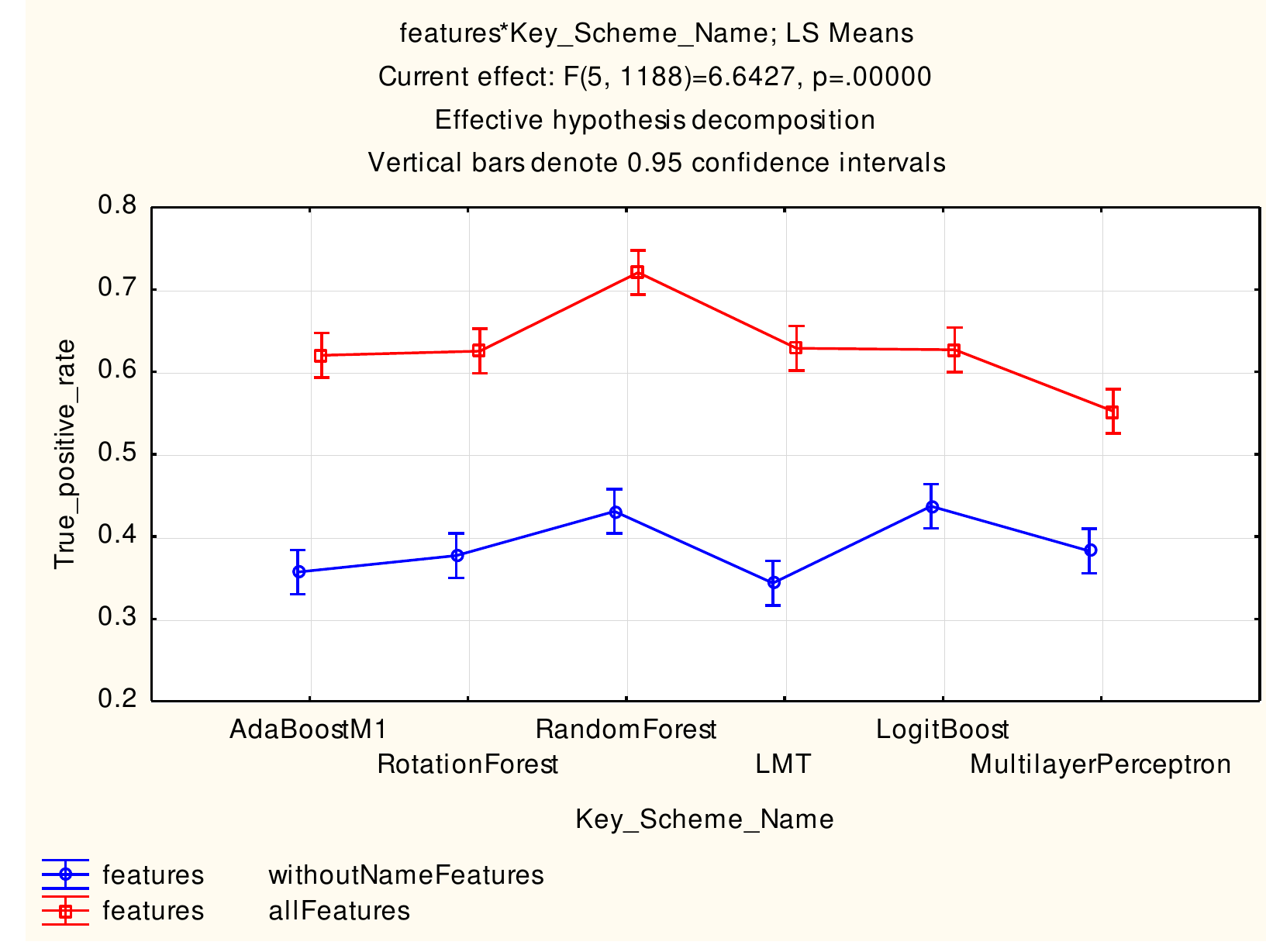}
   \label{fig:de_tpr}
 }%
\end{subfigure}\hfill
 \begin{subfigure}[False Positive Rates Results]{\includegraphics[width=.45\linewidth]{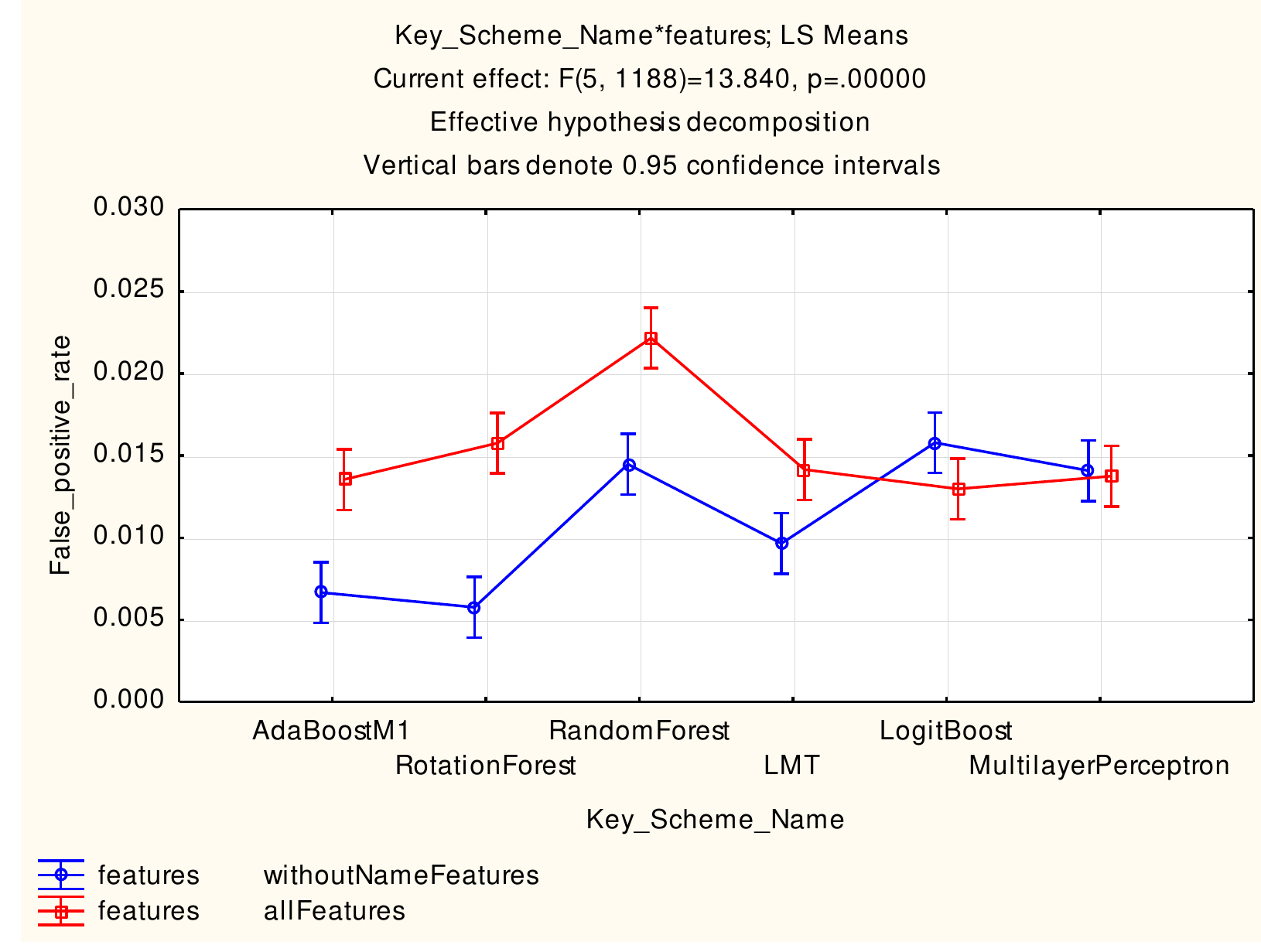}
   \label{fig:de_fpr}
 }%
\end{subfigure}

\caption{De-anonymizing user's identity scenario results (marked in blue) compared to matching users across two social networks scenario  results (marked in red)}
\label{fig:de_results}
\end{figure}

In this scenario, we constructed the classifiers without using names based features like SOUNDEX. The six classifiers' performance results in terms of AUC, accuracy, TPR, and FPR in this scenario are presented in Figure~\ref{fig:de_results}. 

As is portrayed  in Figure~\ref{fig:de_auc}, all of the classifiers' AUC results in this scenario, which are marked in blue, were above $0.8$. The Random Forest classifier demonstrated the best performance and significantly outperformed the other classifiers. 
The null-hypothesis, namely, that all methods' performances were equal and that the observed differences among their AUC values are merely random, was rejected using the ANOVA test, with $F(5,1188) =12.074, p<0.01$. 
Additionally, as it shown in Figure~\ref{fig:de_accuaracy}, in this scenario all six classifiers presented had accuracies of above 93.5\%, whereas unlike the AUC results, the Rotation Forest classifier demonstrated the best performance and significantly outperformed the Multilayer Perceptron and LMT classifiers.  
The null-hypothesis, namely, that all methods' performances were equal and the observed differences among their accuracy values are merely random, was rejected using the ANOVA test, with $F(5,1188) =3.4936, p<0.01$. 
In this scenario the accuracy value was less than that of the results obtained by using all the features (red line), but is nevertheless, still very high. 

Figures~\ref{fig:de_tpr} and~\ref{fig:de_fpr} present the true positive and false positive rates, respectively. It can be seen that unlike the AUC and the accuracy results, the LogitBoost classifier demonstrated the best TPRs and significantly outperformed the AdaBoostM1 and LMT classifiers.  Additionally, 
it can be seen that all of the classifiers' FPRs were under 0.02, where the Rotation Forest classifier demonstrated the best performance and significantly outperformed the other classifiers. In both cases, the null-hypothesis, namely, that all methods' performances were equal and the observed differences among their TPR and FPR values are merely random, was rejected using the ANOVA test, with $F(5,1188) =6.6427, p<0.01$ for the TPR case, and  $F(5,1188) =13.840, p<0.01$ for the FPR case.

\subsection{Feature Analysis Results}
\label{sec:features_results}
To better understand which features were most useful to our classification algorithms in the first scenario, we analyzed the various features' importance using the ``all-but-x'' and ``only-x'' procedures which were described in
Section~\ref{sec:feature_analysis}. 
In both of these procedures, due to its high performance in the matching users across two social networks scenario (see Section~\ref{sec:matching_results}), we selected the LogitBoost algorithm to construct all the entity matching classifiers.

\begin{figure}[ht]
\centering
\includegraphics[width=\linewidth]{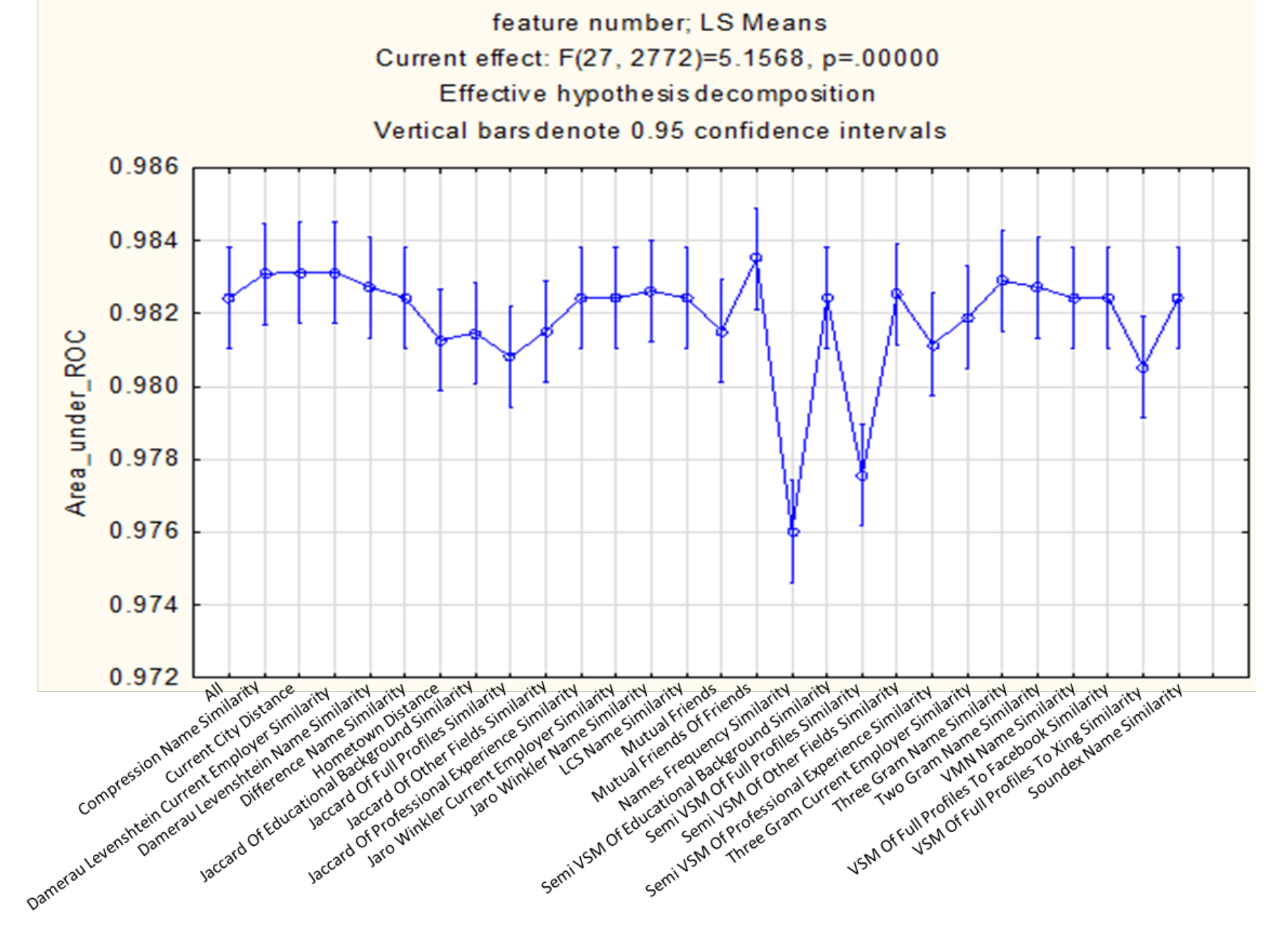}

\caption{AUC performance of ``all-but-x'' features}
\label{fig:all_but_results}
\end{figure}

Figure~\ref{fig:all_but_results} presents the performance of ``all-but-x'' procedures for all the features. As shown, the features in which the dismissal results have the largest drop in the AUC measure are
the Names Frequency Similarity (namesFrequencySimilarity) and 
Semi Vector Space Model (semiVSMOfFullProfilesSimilarity). The null-hypothesis, or the notion that 
the ``all-but-x'' procedure for all features performed the same, and that the observed
differences among the AUC values are merely random was rejected using the ANOVA test,
with $F (27, 2772)=5.1568, p<0.01$. Notice that the Names Frequency Similarity and
the Semi Vector Space Model features belong to two different categories of features and together they calculate the similarity based on all user profile attributes.

\begin{figure}[ht]
\centering
\includegraphics[width=\linewidth]{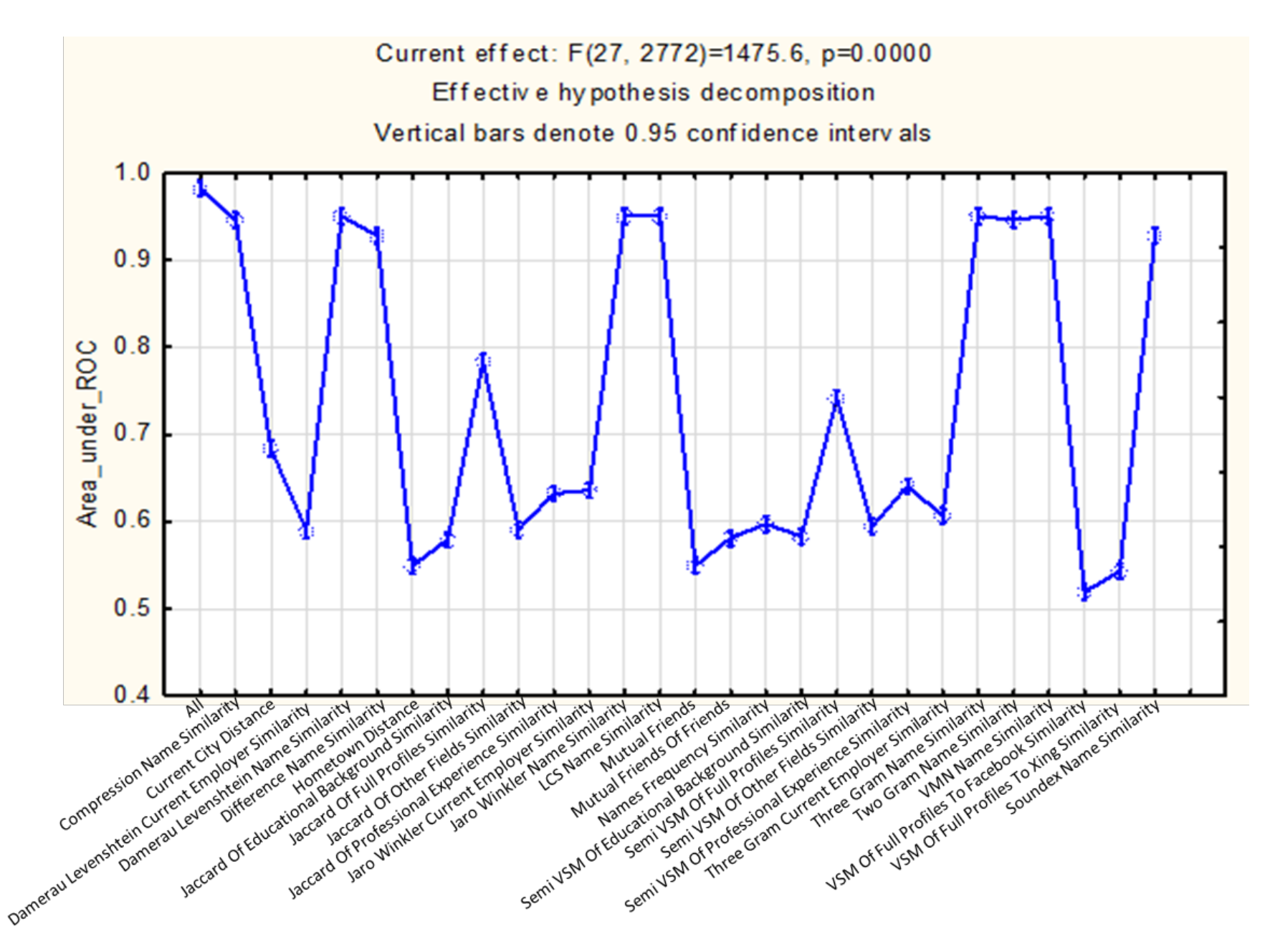}

\caption{AUC performance of ``only-x'' features. }
\label{fig:only_x_results}
\end{figure}

Figure~\ref{fig:only_x_results} presents the performance of the ``only-x'' procedure for all the features. The best single feature for the AUC measure was the Jaro-Winkler Name Similarity (jaroWinklerNameSimilarity) with an AUC of
0.9504. Following it were the Damerau Levenshtein Name Similarity (damerauLevenshteinName) and the LCS Name Similarity (lcsNameSimilarity) with
an AUC of 0.9103, the 3-Gram Name Similarity (threeGramNameSimilarity) and VMN Name Similarity (vmnNameSimilarity) with an AUC of
0.9502. Notice that all five of these features belong to the name based features category. 
The null-hypothesis, the notion that the ``only-x'' procedure for all features performed the same and the
observed differences among its AUC values are merely random, was rejected using the
ANOVA test, with $F(27, 2772) =1475.6, p<0.01$.

\section{Discussion}
\label{sec:disc}
By analyzing the results presented in Section~\ref{sec:results}, the following can be noted.
First, in the first scenario of matching two users across two social networks, among all tested machine learning algorithms, the LogitBoost classifier performed best, with an especially remarkable AUC result of 0.982 (see Figure~\ref{fig:all_auc}), an accuracy of 95.9\% (see Figure~\ref{fig:all_accuaracy}), and a FPR of under 0.014. These results indicate that our proposed algorithms can match users across two social networks with very high accuracy and a low false positive rate.

Secondly, the proposed method for the second entity matching scenario, in which we utilized a user's details in one social network to identify his or her profile with a similar name in another social network, presented relatively good results (see Figure~\ref{fig:search_results1}). In the second scenario, among all tested algorithms, the LogitBoost classifier performed the best with an AUC of 0.88 (see Figure~\ref{fig:search_auc}) and an accuracy of over 80.55\% (see Figure~\ref{fig:search_accuaracy}). This result indicates that if we know the name of the person we are looking for, with an accuracy of 80.5\%, we can find the person's account in a second social network. 

Thirdly, the proposed method for the third entity matching scenario
in which we simulated a case where a person's name and profile details on one social network were given, and we wanted to identify his or her hidden account which contained a pseudonym on different social networks, presented relatively good results (see Figure~\ref{fig:de_results}).
In this scenario the Rotation Forest classifier presented the highest accuracy, over 94.5\% (see Figure~\ref{fig:de_accuaracy}).
This result means that even in the case that two users have completely different names, our algorithm can still infer that these users belong to the same real person. 

Lastly, according to the ``only-x'' AUC results, which were presented in Figure~\ref{fig:only_x_results}, one can see that the top five features were in the name based features category.  This result indicates that the name based features have a strong direct impact on the classification.
Conversely, based on the ``all-but-x'' procedure results, the features whose dismissal resulted in the largest drop in the AUC measure were the Names Frequency Similarity and 
the Semi Vector Space Model  (see Figure~\ref{fig:all_but_results}). These two features belong to two different categories. One belongs to the name based features category and the other one to the general user information-based features category. This indicates that all users' profile fields should indeed be taken into account when possible, and not simply the users' names.

\section{Conclusions and Future Work}
\label{sec:conclusion}
In this study we presented a supervised learning method to match user profiles across multiple OSNs. Our method is based on machine learning techniques that use a variety of features extracted from a user's profile as well as their friends' profiles. Different classifiers were trained and used to rank the probability that two user profiles from different OSNs belong to the same individual. The classifiers utilized multiple features of mainly three types: (a) name based features; (b) general user information-based features; and (c) social network topological based features. 

The proposed method was evaluated using real-life data collected from two OSNs, Facebook and Xing. The proposed method was evaluated on three entity matching scenarios and achieved an AUC of up to 0.982 (see Section~\ref{sec:eval_results}).
This high performance can be attributed to two main factors: the usage of machine learning techniques and the usage of a large variety of features. This high result is evidence that user identification based on web profiles is conceptually and practically possible. 
We can conclude that the LogitBoost algorithm seems to be the most appropriate algorithm to solve our particular problem over all the classification algorithms. 
Although we have seen that the name based features are the most important ones, the combination of all 27 features achieves the best results. Moreover the name based features have a strong direct impact on the classification. Additionally, general user information based features are important features. Therefore, the similarity between two user profiles from two different OSNs should be calculated, not simply using users' names but also using all the users' profile fields.

The proposed method has some limitations. First, we calculated the similarity between two users based on only users' profiles (the ``About'' page) and users' social network topology. However, there are networks such as Twitter in which there is little information about a user and more information about his or her activities in the network (e.g., posts and comments). Perhaps we should also use users' activities in the social network to base the calculation of the similarity. Secondly, name based features have a strong direct impact on the classification. Therefore when two user profiles have different names, it is more difficult to determine if this pair of users represents the same individual. Thirdly, to actually implement this method, enough pairs of members belonging to two different social networks and representing the same entity would be required. Lastly, we only used public profiles to test our algorithms.

In future research, we may extend this process to more than two networks. Moreover, in this research we assumed that a real person has only one user profile page per OSN site. In future work, we may extend this process to find duplicate accounts of a person on the same OSN. In this study we only used the public profile of a user. In future research we may extend this process to include private profiles as well. 

\section*{Acknowledgment}
We would like to thank Jennifer Brill and Marisa Timko for proofreading and editing this article.



\end{document}